\documentclass[draft,jgrga]{agutex}

 \usepackage{lineno}
 \linenumbers*[1]
\usepackage{color}
\usepackage{soul}
\sethlcolor{yellow}
  \usepackage[dvips]{graphicx}
\setkeys{Gin}{draft=false}
\authorrunninghead{TSUCHIYA ET AL.}

\titlerunninghead{GAMMA RAYS FROM WINTER THUNDERCLOUDS}

\authoraddr{H. Tsuchiya,
High-energy Astrophysics Laboratory, Riken, 2-1, Hirosawa, Wako, 
Saitama 351-0198, Japan.
(htsuchiya@riken.jp)}
\authoraddr{T. Enoto,
Kavli Institute for Particle Astrophysics and Cosmology, 
Department of Physics and SLAC National Accelerator Laboratory, 
Stanford University, Stanford, CA 94305, USA
(enoto@stanford.edu)}
\authoraddr{S. Yamada,
Department of Physics, University of Tokyo, 7-3-1, Hongo, 
Bunkyo-ku, Tokyo 113-0033, Japan.
(yamada@juno.phys.s.u-tokyo.ac.jp)}
\authoraddr{T. Yuasa,
Department of Physics, University of Tokyo, 7-3-1, Hongo, 
Bunkyo-ku, Tokyo 113-0033, Japan.
(yuasa@juno.phys.s.u-tokyo.ac.jp)}
\authoraddr{K. Nakazawa,
Department of Physics, University of Tokyo, 7-3-1, Hongo, 
Bunkyo-ku, Tokyo 113-0033, Japan.
(nakazawa@phys.s.u-tokyo.ac.jp)}
\authoraddr{T. Kitaguchi,
 Division of Physics, Mathematics, and Astronomy,
 California Institute of Technology,
 1200 East California Boulevard,
 Pasadena, CA 91125, USA
(kitaguti@caltech.edu)}
\authoraddr{M. Kawaharada,
Department of High Energy Astrophysics,
Institute of Space and Astronautical Science, JAXA, 3-1-1,
Chuo-ku, Sagamihara, Kanagawa 252-5210, Japan.
(kawahard@astro.isas.jaxa.jp)}
\authoraddr{M. Kokubun,
Department of High Energy Astrophysics,
Institute of Space and Astronautical Science, JAXA, 3-1-1,
Chuo-ku, Sagamihara, Kanagawa 252-5210, Japan.
(kokubun@astro.isas.jaxa.jp)}
\authoraddr{H. Kato,
High-energy Astrophysics Laboratory, Riken, 2-1, Hirosawa, Wako, 
Saitama 351-0198, Japan.
(hkato@riken.jp)}
\authoraddr{M. Okano,
High-energy Astrophysics Laboratory, Riken, 2-1, Hirosawa, Wako, 
Saitama 351-0198, Japan.
(mokano@crab.riken.jp)}
\authoraddr{K. Makishima,
Department of Physics, University of Tokyo, 7-3-1, Hongo, 
Bunkyo-ku, Tokyo 113-0033, Japan.
(maxima@phys.s.u-tokyo.ac.jp)}
\begin{document}
%
%

\title{Long-duration gamma-ray emissions from 2007 and 2008 winter thunderstorms}
%
%






 \authors{H. Tsuchiya, \altaffilmark{1}
 T. Enoto, \altaffilmark{2}
 S. Yamada, \altaffilmark{3}
 T. Yuasa, \altaffilmark{3}
 K. Nakazawa,\altaffilmark{3}
 T. Kitaguchi,\altaffilmark{4}
 M. Kawaharada,\altaffilmark{5}
 M. Kokubun,\altaffilmark{5}
 H. Kato,\altaffilmark{1}
 M. Okano,\altaffilmark{1}
and  K. Makishima\altaffilmark{3}
 }

%
%


\begin{abstract}
The Gamma-Ray Observation of Winter THunderclouds (GROWTH) experiment, 
consisting of two radiation-detection subsystems, has been operating since 2006 on the premises of Kashiwazaki-Kariwa nuclear 
power plant located at the coastal area of Japan Sea. 
By 2010 February, GROWTH 
detected 7 long-duration $\gamma$-rays emissions associated with winter thunderstorms.
Of them,  two events, obtained on 2007 December 13 and 2008 December 25, are reported.
On both occasions, all inorganic scintillators (NaI, CsI, and BGO) of the two subsystems
detected significant $\gamma$-ray signals lasting for $>$1 minute.  
Neither of these two events were associated with any lightning. In both 
cases, the $\gamma$-ray energy spectra extend to 10 MeV, 
suggesting that the detected $\gamma$-rays are produced by 
relativistic electrons via bremsstrahlung.  
Assuming that the initial photon spectrum at the source is expressed by a power-law function, 
the observed photons can be interpreted as being radiated from a 
source located at a distance of 290 $-$  560 m for the 2007 event and 110  $-$ 690 m 
for the 2008 one, both at 90\% confidence level.
Employing these photon spectra, the number of relativistic electrons
is estimated as $10^9 - 10^{11}$.
The estimation generally agrees with those calculated
based on the relativistic runaway electron avalanche model.
A GROWTH photon spectrum,  summed over 3 individual events including 
the present two events and another reported previously, has similar 
features including a cut-off  energy,  to an averaged spectrum of
terrestrial gamma-ray flashes.
\end{abstract}

%
%

\begin{article}

%
%

\section{Introduction}
Nonthermal X-ray and $\gamma$ ray emission, typically lasting for a few seconds to $\sim10$ minutes, 
has been observed from thunderstorm 
activity, with detectors on board an airplane~\citep{MP_1985} and a ballon~\citep{Eack_1996,Eack_2000},
high-mountain detectors~\citep{SUSZ_1996,EAS_2000,Chub_2000,Alex_2002,Norikura_2004,MtFuji_2009,norikura_2009, Chili_2010},
and ground-based ones~\citep{MONJU_2002,growth_2007}.
Interestingly, they do not appear to clearly coincide with lightning processes such as
stepped leaders or return strokes. In contrast, much shorter energetic radiation bursts, lasting only for tens of milliseconds or less,  
are often associated with lightning discharges. Though not necessarily homogeneous, they 
include terrestrial gamma-ray flashes (TGFs)~\citep{BATSE,RHESSI,RHESSI_2009,Fermi_2010a,Fermi_2010b,AGILE_MCAL,AGILE_MCAL_PRL}, 
natural lightning~\citep{moore_2001,Dwyer_natural_2005,Howard_2008,Yoshida_2008,Chub_2009},
and rocket-trigerred ones~\citep{Dwyer_sci_2003,Dwyer_exp_typspe,Dwyer_10MeV_2004}.

In this way, it has recently become clear that apparently 
two types of radiation bursts with distinct duration are associated with thunderstorm activity. 
Although it is uncertain whether or not these two types have a common source mechanism, 
recent observations as well as theoretical works generally suggest that these bursts, especially 
short-duration ones, are produced by processes involving acceleration and multiplication of a 
background population of electrons.

Various numerical kinetic calculations~\citep{RREA_1992,Roussel_1994,Bell_1995,Le_1996,RREA_1997,Mili_1999,RREA_2007,Roussel_SSR_2008} 
and Monte Carlo simulations~\citep{Le_1999,Dwyer_MC2003, Babich_2005,Babich_MC2007}
commonly indicate that most of prompt nonthermal photons from lightning 
discharges are radiated, via bremsstrahlung, by relativistic electrons, which in turn are  produced through 
mechanism involving relativistic runaway electron avalanche (RREA):
some seed electrons, produced by e.g. cosmic rays, can be accelerated into relativistic regime if they can 
gain energies from the high electric fields in thunderclouds 
fast enough to overcome their total energy losses, due mainly to ionization.
Then, they collide with air molecules and ionize them.
Some of the faster newborn secondary electrons are also accelerated to higher energies, hence increasing in 
their number. Finally, they will emit a detectable flux of nonthermal photons via 
bremsstrahlung. 

Early observations of long-duration bursts, though limited in number,  measured X-ray fluxes in a few keV to a few 
hundred keV range, or $\gamma$-ray fluxes in MeV regions, suggesting 
that these prolonged emissions are also due to relativistic electrons~\citep{MP_1985,Eack_1996,Eack_2000,EAS_2000,Chub_2000}.
Several recent observations~\citep{growth_2007,MtFuji_2009,norikura_2009,Chili_2010}
have reinforced the suggestion, by detecting photon spectra extending clearly to 10 MeV or higher, and have
given evidence that those long-duration $\gamma$ rays are also produced via bremsstrahlung.
These results naturally lead to a view that long-duration events are also caused by relativistic runaway electrons.
However, compared with short-duration ones, the nature of long-duration bursts
have remained less understood, due primary to the lack of a sufficiently large sample.
For example, it is still unclear how the electron acceleration process keeps
operating for such long durations.
In addition, the relation between short-duration bursts and long-duration ones is unknown.

Aiming at detections of radiation bursts from thunderstorm activity, we have been 
operating the Gamma-Ray Observations of Winter THunderclouds (GROWTH) experiment 
since 2006 December 20.
In this paper, we report on successful GROWTH detections of two long-duration $\gamma$-ray bursts extending to 10 MeV.
Using the acquired $\gamma$-ray data, the source distance, its spatial extent, and
the number of relativistic electrons involved therein are estimated. Then,  
a $\gamma$-ray spectrum which sums up 3 GROWTH detections 
is compared with cumulative TGF spectra obtained by two independent space observations. Based on
these results, we quantitatively discuss the production mechanism of prolonged $\gamma$-ray bursts 
from winter thunderclouds.  

\section{The GROWTH Experiment}\label{exp}%
The GROWTH experiment, comprising two independent subsystems, has been operating successfully  at a roof of
a building of Kashiwazaki-Kariwa nuclear power plant in Niigata Prefecture, Japan. Figure~\ref{fig:location} shows the 
location of the plant, facing the Japan Sea, and the GROWTH experimental site therein. The geographical longitude, latitude, 
and altitude of the experimental site are $138^\circ36^{'}$E, $37^\circ26^{'}$N, and 40 m above sea level, respectively.
This coastal area is frequently struck by strong thunderstorms in winter seasons. 
Actually, before the GROWTH experiment started working, radiation monitors (filled circles in 
Fig.~\ref{fig:location}), which are arranged at around 300 m $-$ 400 m intervals in the plant,
occasionally observed $ >3$ MeV intense radiation enhancements in winter seasons, which 
are difficult to ascribe to so-called radon washouts because these rainfall-related episodes would  mainly 
cause increases at $<3$ MeV energies~\citep[e.g.][]{Yoshioka_1992,Yamazaki_2002}.
Each radiation monitor consists of a $\phi 5.1 \, \mathrm{cm}\times 5.1\, \mathrm{cm}$ NaI (Tl) scintillation counter, and
a spherical ion chamber with a volume of $\sim14$ L that contains Ar gas. The former covers the 50 keV $-$ 3 MeV energy range, 
while the later operates in $>$ 50 keV. 
However, the radiation monitors have too poor a time resolution of 30 sec, together with the too limited energy bands,  
to understand the nature of those phenomena.
The GROWTH experiment is expected to provide much improved knowledge on these sporadic events.

The pictures and drawings of the two subsystems are given in \citet{Enoto_30thicrr} and \citet{growth_2007}. 
One of them (Detector-A) uses two cylindrical NaI (Tl) scintillators (density $= 3.67\, \mathrm{gcm^{-3}}$),
having a diameter and a height of both 7.62 cm. 
In order to actively shield them from natural low-energy ($< 3$ MeV) environmental radiation
(e.g. from $^{40}\mathrm{K}$), the NaI scintillators are individually 
surrounded by well-shaped BGO ($\mathrm{Bi_{4}Ge_{3}O_{12}}$; density $= 7.1\, \mathrm{gcm^{-3}}$) scintillators,
with the thickness on the side and bottom being 1.27 cm and 2.54 cm, respectively.
The BGO scintillators geometrically shield the central NaI up to a solid angle of $2.4\pi$ str, or $0.6\times 4\pi$.
Thus, the NaI scintillators have a higher sensitivity toward the sky direction. 
The two central NaI scintillators and the BGO shields are operated over an energy range of 40 keV $-$ 10 MeV.
Output signals from photomultiplier tubes, attached to the NaI and BGO, are fed 
individually to a 12 bit 8ch VME-analog-to-digital converter [ADC (CP 1113A)] with a time 
resolution of 10 $\mu$sec, and is recorded on event-by-event basis.

As another feature of Detector-A, a 0.5 cm thick plastic scintillator with an area 
of $30.5 \,\mathrm{cm}\times 15.2 \,\mathrm{cm} = 464 \,\,\mathrm{cm^2}$ is placed above the two 
NaI scintillators, and operated with a threshold energy of $>1$ MeV.
It has a high detection efficiency for charged particles, 
while it is almost transparent to photons due to its thinness and higher threshold. 
Utilizing this feature, we can separate charged particles from photons, and efficiently exclude 
background cosmic-ray muons, which typically deposit $>1$ MeV energies, from events in the two NaI scintillators.
Specifically, an event in either of the two NaI scintillators is judged as a
charged particle if it give a simultaneous hit (with 10 $\mu$ sec) in the plastic scintillator.
Thus, utilizing signals of the BGO and plastic scintillators both in anticoincidence, 
the central NaI scintillators effectively detect photons, generally arriving from a sky direction.

Aiming at an independent radiation measurement, another subsystem (Detector-B) 
was installed $\sim 10$ m apart from Detector-A. It consists
of spherical NaI (Tl) and  CsI (Tl) scintillators (density $= 4.51\, \mathrm{gcm^{-3}}$), both with a diameter of
7.62 cm. The former operates in 40 keV $-$10 MeV, 
while the latter covers a higher energy range of 300 keV $-$ 80 MeV.
Unlike Detector-A, these scintillators have 
omni-directional sensitivity because they have no shields such as the BGO
or plastic scintillators.
Output signals of two photomultiplier tubes, attached to the NaI and CsI crystals, 
are sampled by a self-triggering electronics system with a 12 bit ADC (AD 574). These events
are accumulated into an ADC histogram, which is recorded every 6 sec. 

Energy calibrations of Detector-A and Detector-B
were carried out, using natural environmental $\gamma$-ray lines of ${}^{214}\mathrm{Pb}$ (0.352 MeV), 
${}^{214}\mathrm{Bi}$ (0.609 MeV), ${}^{40}\mathrm{K}$ (1.46 MeV), and $^{208}\mathrm{Tl}$ (2.61 MeV). 
Then,  especially for the CsI of Detector-B,  cosmic-ray muons,  giving energy deposits with its peak of 
around 35 MeV,  were also utilized.  Basically, these calibrations are performed and checked 
by an offline analysis.

In addition to those radiation detectors,  the GROWTH system utilizes three optical sensors and 
an electric-field mill as environmental monitors.  
Each optical sensor consists of a hand-made analog circuit, and a silicon 
photodiode (HAMAMATSU S1226-8BK) which is sensitive over a wavelength range of 
320 nm $-$ 1000 nm (with its peak at 750 nm). They measure
environmental visible light in coarsely different directions; sea side, zenith direction, 
and anti-sea side.
The output signals are fed to a 12 bit VME-ADC, and recorded  every 0.1 sec. The electric field mill
is a commercial product (BOLTEK EFM-100). 
Its analog output is fed to
a 12 bit ADC (AD 7892), and recored as electric-field strength 
between $\pm 100\, \mathrm{kV\, m^{-1}}$, with a resolution of 50 $\mathrm{Vm^{-1}}$.
\section{Results}\label{res}%
\subsection{Count histories of the inorganic scintillators}\label{res:count}%
\begin{figure}
\noindent\includegraphics[scale=0.7]{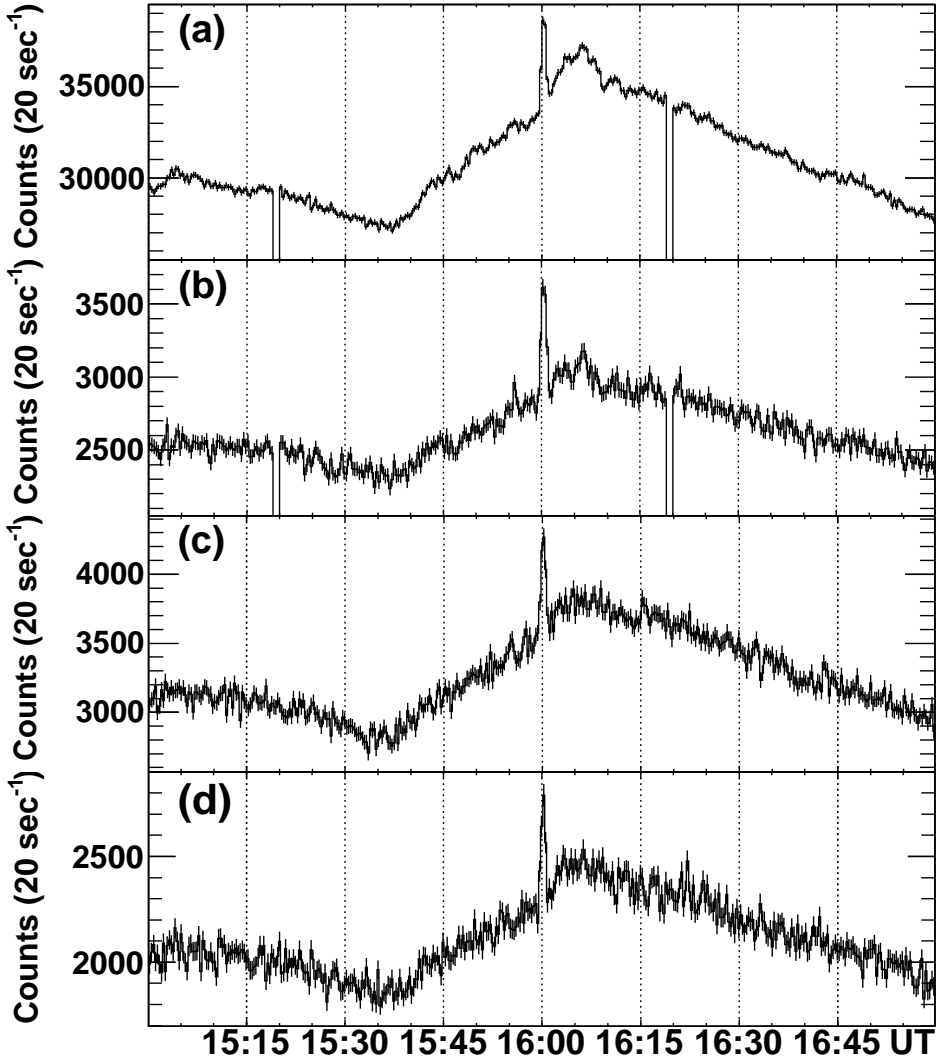}
 \caption{Count rates per 20 sec of the 4 inorganic scintillators over 15:00 $-$ 17:00 UT on 2007 December 13.
 Panels (a) and (b) show the $>$ 40 keV count rates from the BGO and NaI scintillators without the anticoincidence
 of Detector-A, respectively,   while panels (c) and (d) represent those of NaI ($>40$ keV) and CsI ($>300$ keV) scintillators of Detector-B, respectively. 
Horizontal axis shows universal time. Error bars are statistical 1$\sigma$. The gaps in panels (a) and (b) are 
due to regular interruptions of data acquisition of Detector-A every hour. }
\label{fig:4inorg_scinti_15_17_lc}
 \end{figure}

\begin{figure}
\noindent\includegraphics[scale=0.7]{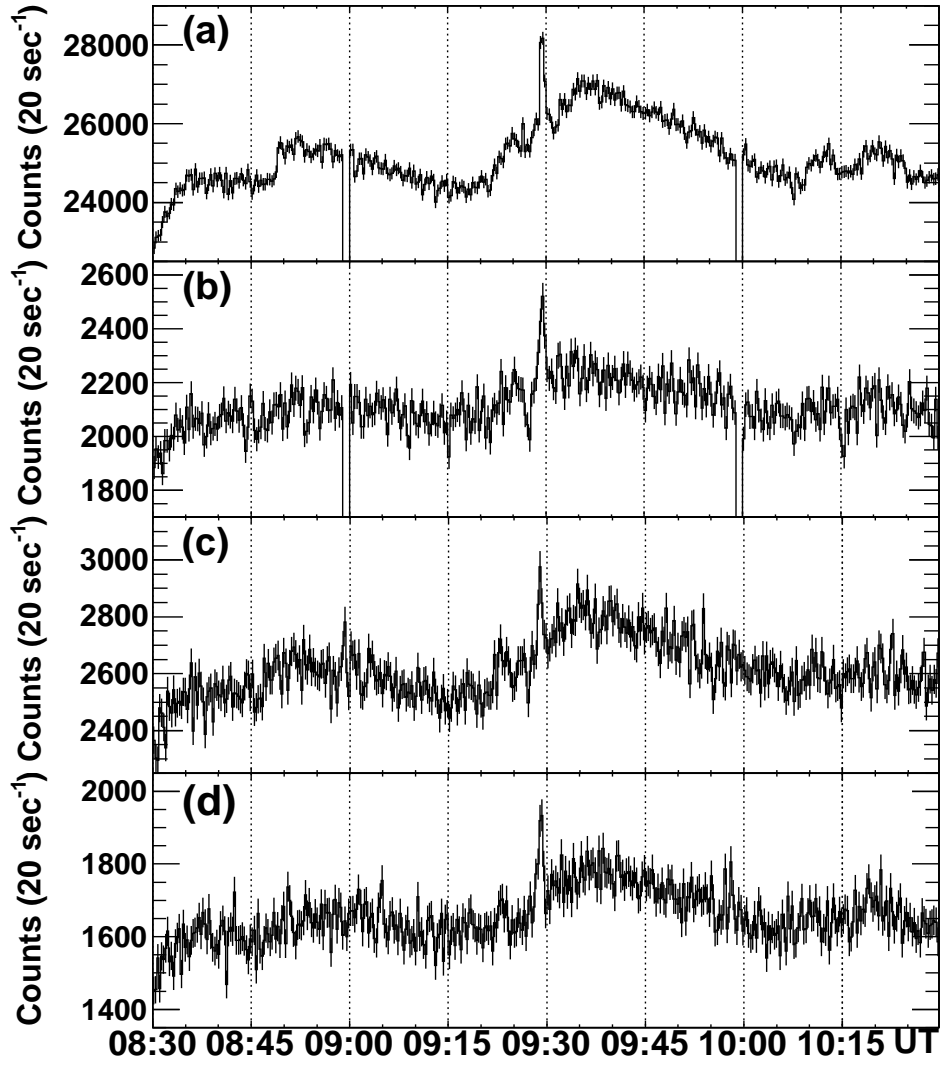}
 \caption{The same as Fig.~\ref{fig:4inorg_scinti_15_17_lc}, but over 8:30 $-$ 10:30 UT on 2008 December 25.}
\label{fig:4inorg_scinti_20081225_lc}
 \end{figure}

Figure~\ref{fig:4inorg_scinti_15_17_lc} shows 
count histories of the 4 inorganic scintillators of Detector-A and B, obtained over 15:00 $-$ 17:00 UT
on 2007 December 13 which corresponds to local midnight (0:00 $-$ 2:00 JST on 2007 December 14). 
For reference, typical background rates per 20 sec, corresponding to
the panels (a), (b), (c), and (d) of Fig.~\ref{fig:4inorg_scinti_15_17_lc}, are 22000, 1700, 2100, 
and 1500,  respectively.
Similarly, Figure~\ref{fig:4inorg_scinti_20081225_lc} gives those over 8:30 $-$ 10:30 UT on 2008 December 25
(17:30 $-$ 19:30 JST on the same day, or local evening). 
On both these days, a strong low pressure system (with $\sim 990$ hPa on the ground) developed over Japan, 
causing thunderstorms at the coastal area of Japan Sea. 
A gradual count increase, followed by a gradual count decrease, generally shows that they are 
due mainly to radioactive radon and its decay products in rain,with their half-lives being 20 $-$ 30 min. 
These effects originating from radionuclides 
are closely investigated by \citet{SUSZ_1996} and \citet{Yamazaki_2002}. 

Superimposed on such gradual count increases, 
a sharp count enhancement is found in all the inorganic scintillators at around 16:00 UT in Fig.~\ref{fig:4inorg_scinti_15_17_lc}, and
at around 9:30 UT in Fig.~\ref{fig:4inorg_scinti_20081225_lc}. Hereafter, we call the former and the 
latter events 071213 and 081225, respectively.
These enhancements, both lasting for $70-80$ sec, are quite different from the radon effects 
and from short radiation bursts associated with lightning discharges.
Among those inorganic scintillators, BGO of Detector-A gave statistically 
the most significant burst detection on both occasions; $30\sigma$ for 071213 and
$19\sigma$ for 081225. This is because it has a higher density and a larger
effective atomic number, and hence a higher stopping power, especially for X/$\gamma$ rays, than the 
other inorganic scintillators used in our system.

\begin{figure}
\noindent\includegraphics[scale=0.45]{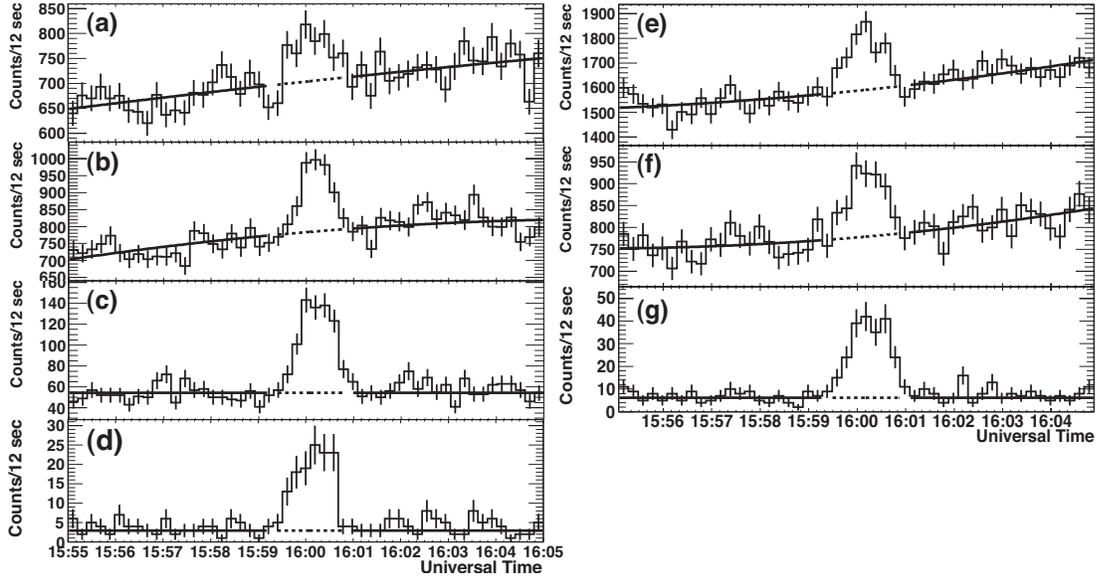}
 \caption{
 Count histories per 12 sec of 071213 in three energy bands from Detector-A 
 (left; summed over the two NaI) and  Detector-B (right; NaI),
 obtained for 15:55 $-$ 16:05 UT. Panels (a), (b) and (c) correspond to 
 $0.04-0.3$ MeV, $0.3-3$ MeV and $>3$ MeV energy bands
 without anticoincidence, respectively.  Panel (d) indicates 
the $>3$ MeV energy band of the NaI with anticoincidence.
Panels (e), (f) and (g) are the same as Panels (a), (b) and (c), respectively, but for Detector-B.
 Solid curves outside the burst period show the estimated 
 background level (see text),  while dashed ones denote the interpolated background level 
 over the burst period. Abscissa represents universal time. Each error bar is statistical $1\sigma$. }
 \label{fig:detab_3ebands_071213}
 \end{figure}

\begin{figure}
\noindent\includegraphics[scale=0.45]{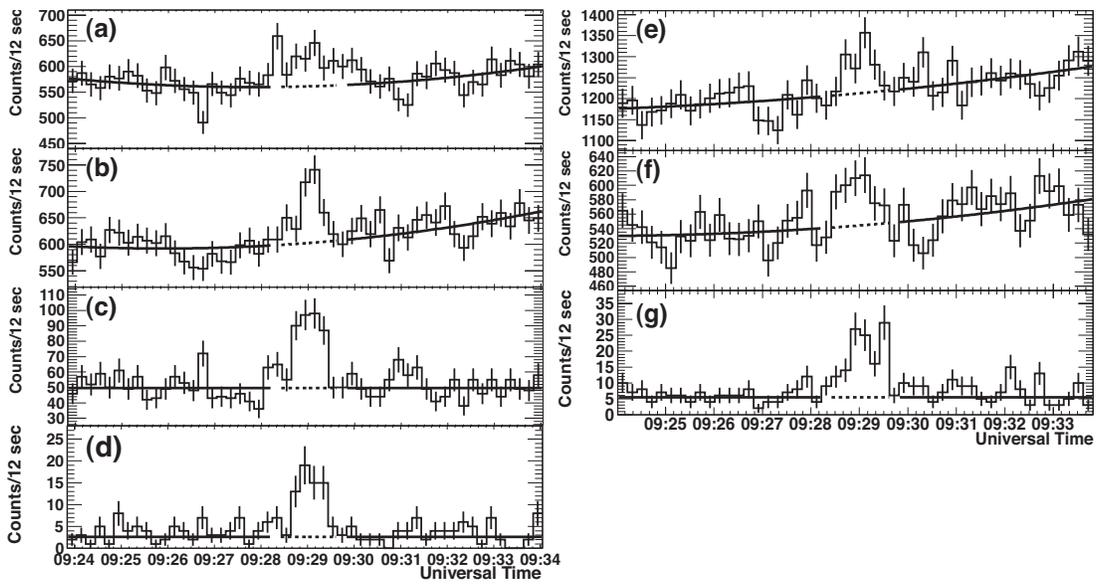}
 \caption{
 The same as Fig.~\ref{fig:detab_3ebands_071213}, but for 081225, 
 obtained over 9:24 $-$ 9:34 UT.
 }
 \label{fig:detab_3ebands_081225}
 \end{figure}

Figure~\ref{fig:detab_3ebands_071213} shows NaI count histories of 071213 in 3 energy bands
from Detector-A and B, while Figure~\ref{fig:detab_3ebands_081225} represents those 
of 081225. For comparison with Detector-B, the data of Detector-A [panels (a), (b) and (c) of Fig.~\ref{fig:detab_3ebands_071213} 
and Fig.~\ref{fig:detab_3ebands_081225}] are presented without the BGO or plastic anticoincidence. 
With a criterion that both the NaI and CsI scintillators of Detector-B simultaneously 
record 10 or higher counts per 12 sec in the $3 -10$ MeV energy band, 
we define burst periods of 071213 and 081225 as
84 sec, 15:59:29 $-$ 16:00:53 UT, and 72 sec, 9:28:29 $-$ 9:29:37 UT, respectively.
For reference, this energy band of either scintillator typically records $\sim4-5$ events per 12 sec
in quiescent periods; so the above criterion (again, not either but both scintillators 
have 10 or higher counts) means approximately $\ge 3.2 - 4.0 \sigma$  above the background.

In order to estimate background levels of individual energy 
bands of Detector-A and B, we excluded data over the burst period (as defined above) 
and the adjacent 12-sec periods. The remaining data in the two lower-energy bands were fitted by
a quadratic function (via $\chi^2$ evaluation), while those in the highest-energy band with a constant.
Table~\ref{tab:netcounts_flux} summarizes the net count increases, obtained by subtracting interpolated background 
(dashed curves of Fig.~\ref{fig:detab_3ebands_071213} and Fig.~\ref{fig:detab_3ebands_081225}) from the total counts 
in the burst period.  Thus, the burst detection is statistically significant in each of the three energy bands on both occasions. 
Table~\ref{tab:netcounts_flux} also gives the observed photon number fluxes above the detectors 
using power-law spectra obtained later (Sec.\ref{res:spectra_modelfit}) and the detector responses of
Detector B derived from a Monte Carlo simulation based on GEANT4~\citep{GEANT4}.  Here, 
the MC simulation was evaluated with radionuclide sources of  $^{60}$Co and $^{137}$Cs.

\subsection{Arrival directions}\label{res:direction}%
Shown in Fig.~\ref{fig:detab_3ebands_071213} and Fig.~\ref{fig:detab_3ebands_081225} are the NaI count rates of 
Detector A at $>3$ MeV energies of 071213 and 081225,  respectively, with [panel (d) in both figures]
and without [panel (c)] anticoincidence.
In both figures, the anticoincidence, which utilizes BGO and plastic signals in logical "OR", is seen to
reduce the NaI background level (solid curves) to $\sim0.05$ times that without anticoincidence. 
In contrast, the NaI-detected burst signal rate decreases due to the anticoincidence only 
to $0.25\pm0.03$ and $0.31\pm0.06$ times the raw rates, for 071213 and 081225, respectively. 
Thus, the burst photons survive the anti-coincidence with 5 $-$ 6 times higher efficiency than the background events.
Similarly, the ratio
of the $>40$ keV NaI [Fig.~\ref{fig:4inorg_scinti_15_17_lc} (b) and Fig.~\ref{fig:4inorg_scinti_20081225_lc} (b)] 
to the $>40$ keV BGO count rates [Fig.~\ref{fig:4inorg_scinti_15_17_lc} (a) and Fig.~\ref{fig:4inorg_scinti_20081225_lc} (a)],
which is normally $\sim$ 0.08 due mostly to  environmental radioactivity coming from omni-directions, 
increased to $0.18\pm0.02$ for 071213 and $0.14 \pm 0.02 $ for 081225. 

The above properties revealed by applying the anticoincidence are 
thought to reflect arrival directions of the 
burst signals. If they came mainly from horizontal or ground directions, 
the anticoincidence on/off ratio and the NaI/BGO ratio would both fall below their normal values, 
because, e.g., 40 keV or 3 MeV $\gamma$ rays horizontally entering Detector-A would be
almost fully or partially (at least 30\%) absorbed/scattered by BGO via photoelectric absorption and Compton scattering.
Accordingly, we conclude that the burst signals arrived from sky directions, not from horizontal or ground directions.
A more quantitative study of arrival directions, employing Monte Carlo simulations, will be reported elsewhere.

\subsection{Burst components}\label{res:comp}%
\begin{figure}
\noindent\includegraphics[scale=0.45]{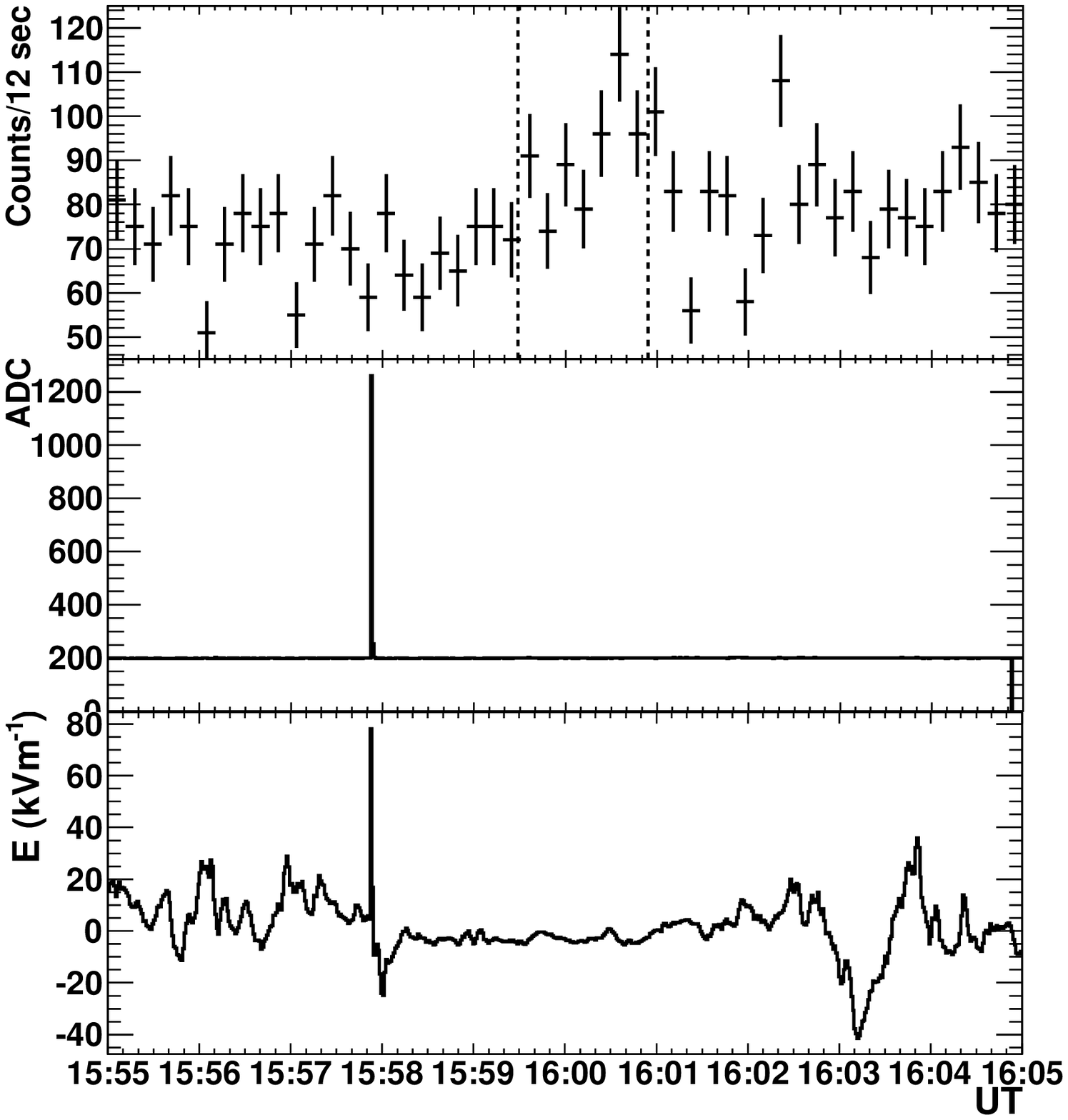}
\hspace{-1.2cm}
\noindent\includegraphics[scale=0.45]{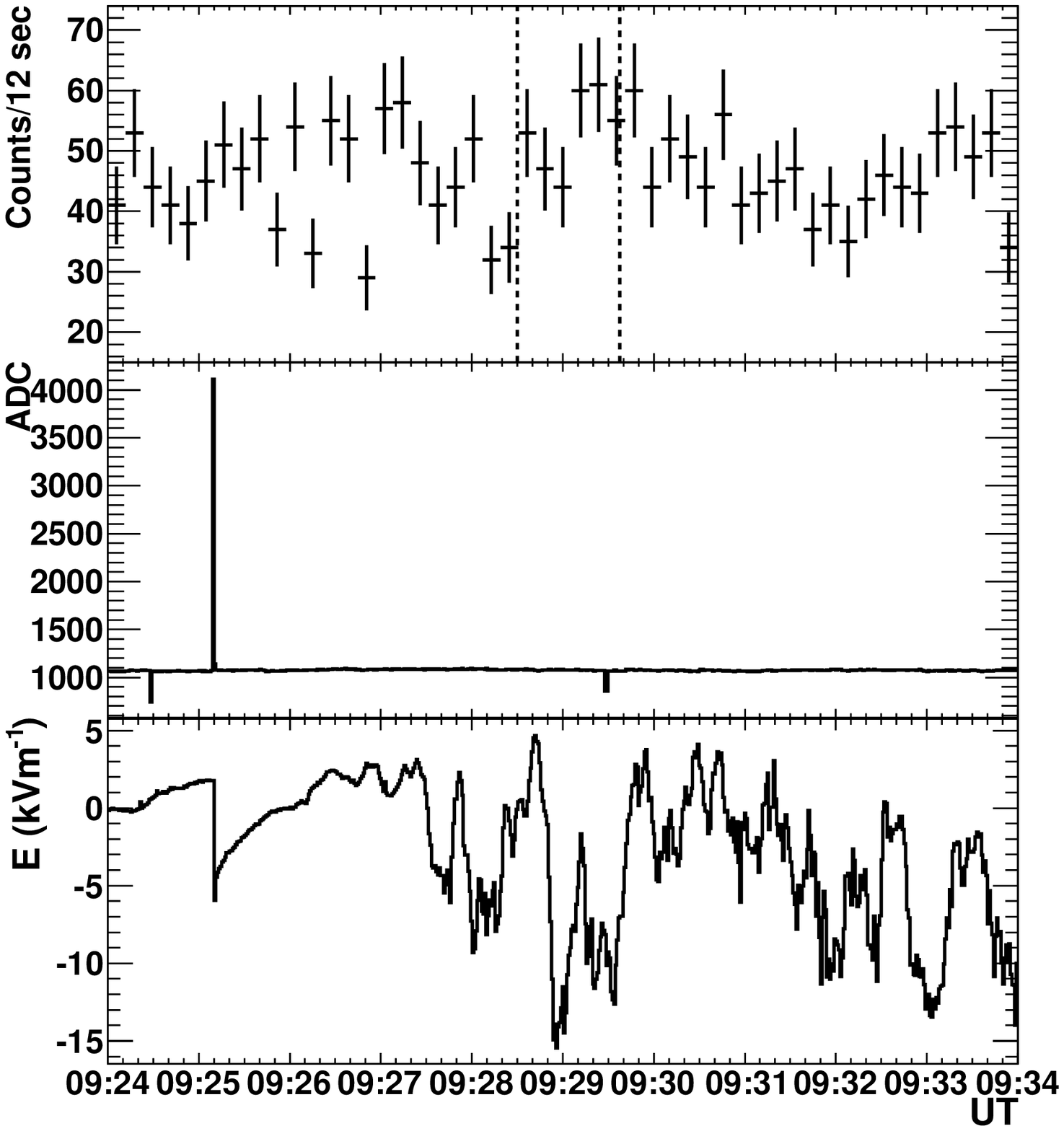}
 \caption{%
 The count-rate histories of the plastic scintillator of Detector A and  the 
 environmental sensors. Left panels show 071213,  obtained over 
 15:55 $-$ 16:05 UT, while right ones represent 081225, obtained 
 for 9:24 $-$ 9:34 UT. Top, middle and bottom panels in both sides represent a $>1$ MeV
 counts every 12 sec from the plastic scintillator, 1-sec optical data variations, 
 and 1-sec electric field ones, respectively.
 All abscissa are universal time. Vertical lines in top panels represent the burst periods.
  }
 \label{fig:pl_env_071213_081225}
 \end{figure}

Figure~\ref{fig:pl_env_071213_081225} shows
count histories of the plastic scintillator ($>1$ MeV) and the environmental sensors.
In coincidence with the apparent signals detected by the inorganic scintillators, 
the 0.5 cm thick plastic scintillator gave count increases in individual burst periods
by $N_\mathrm{pl}=$ $160\pm30\, (5.3\sigma)$ for 071213 and $72 \pm 18 \,(4\sigma)$ for 081225
[top panels of Fig.~\ref{fig:pl_env_071213_081225}]. 
Presumably these plastic signals are composed of $\gamma$ rays and charged particles,
most likely electrons, which are either accelerated primaries or secondary ones produced by high-energy photons 
via Compton scattering.
Below, we estimate how $\gamma$ rays and electrons
contribute to $N_\mathrm{pl}$, and estimate the electron flux above 1 MeV.

First, a Monte Carlo simulation using GEANT4 predicts that the plastic scintillator 
has a low detection efficiency, $0.5 - 1\%$, for $> 1$ MeV $\gamma$ rays, 
while that  for $>1$ MeV electrons reaches $75 - 90\%$. Next, using power-law spectra obtained later (Sec.\ref{res:spectra_modelfit}) 
and the effective area of the NaI scintillator of Detector B yield $1-10 $ MeV photon number fluxes above the GROWTH system 
as $(17.9 \pm 1.8)\times10^{-2}\,\, \mathrm{cm^{-2}s^{-1}}$ for 071213,
and $(9.2\pm1.8) \times 10^{-2}\,\, \mathrm{cm^{-2}s^{-1}}$ for 081213. Then, 
multiplying these fluxes by the GEANT4-derived detection efficiency for $\gamma$ rays of the 
plastic scintillator and the area of the plastic scintillator, $464\,\, \mathrm{cm^2}$, 
$\gamma$-ray produced counts contributing to $N_\mathrm{pl}$
are estimated as  $N_\gamma = 70\pm7$ for 071213, and $31\pm6$ for 081225. 
Finally, subtracting $N_\gamma$ from $N_\mathrm{pl}$, the contribution of electrons 
is obtained as $90 \pm 30$ for 071213, and $41\pm19$ for 081225. 
Although these numbers, when taken at their face values, imply a significant electron
contribution to $N_\mathrm{pl}$, here we conservatively regard them as upper limits.
Then, a 95\% confidence level upper limit on the electron flux above 1 MeV of 071213 and 081225 
is computed as $0.5 \times 10^{-2}\,\, \mathrm{cm^{-2}s^{-1}}$  
and $0.3 \times10^{-2} \,\, \mathrm{cm^{-2}s^{-1}}$, respectively. These upper limits are more than an order of magnitude lower
than the $1-10$ MeV $\gamma$-ray fluxes. Therefore, 
the observed burst signals arriving at the GROWTH system are inferred to be dominated by photons, rather than electrons.

\subsection{Comparison with signals from environmental sensors}\label{res:env}
The visible-light sensor (middle panels of Fig.~\ref{fig:pl_env_071213_081225}) 
recorded an extremely intense signal, lasting  $\le 1$ sec,  at 15:57:50 UT and 9:25:09 UT, for 071213 and 081225, respectively.
In coincidence with the recorded optical flashes, the electric field rapidly changed its polarity from positive to 
negative (bottom panels of Fig.~\ref{fig:pl_env_071213_081225}). 
These indicate that a lightning discharge occurred. 
However, their occurrence is  well separated from the $\gamma$-ray bursts themselves, namely,
100 sec and 180 sec prior to the 071213 and 082125 commencements, respectively. 
Thus, we conclude that neither of the present two $\gamma$-ray bursts coincided 
with lightning discharges.

Prior to the present work, \citet{growth_2007, norikura_2009} have reported similar lack of coincidence between prolonged $\gamma$-ray bursts and lightning 
discharges. In \citet{growth_2007}, a long-duration burst, lasting 40 sec, was detected by the GROWHT system 70 sec prior to lightning,   
while in \citet{norikura_2009}, no lightning discharges were measured over 5 minutes before or after a prolonged ($\sim90$ sec) burst detected
at a high-mountain detector. 
These previously reported events, observed during thunderstorms, have also been considered to be associated with thunderclouds. 

In the same manner, we associate the present bursts with the thunderclouds, rather than to lightning discharges.
Actually, rainfall-thunder observation data\footnote{raifall-thunder observation data are available from http://thunder.tepco.co.jp/}, 
provided by a laser observation system operated by Tokyo Electric Power Company, showed that 
thunderclouds approached the Kashiwazaki-Kariwa nuclear plant from the sea side on both occasions, and passed over it 
during the 10 minutes. 

\subsection{Energy spectra}\label{res:spectra_modelfit}%
\begin{figure}
\noindent\includegraphics[scale=0.50]{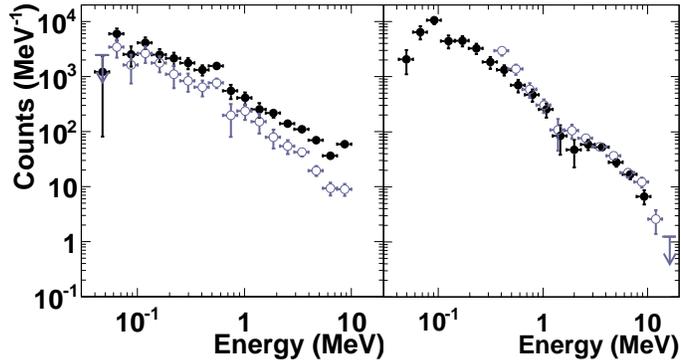}
 \caption{%
 Background-subtracted spectra of Detector-A (left) and B (right) of 071213.
 Black and gray points in the left panel indicate the NaI data without and with anticoincidence,
 respectively, while those in the right panel show the NaI and CsI scintillators, respectively.
 All error bars quoted are statistical 1$\sigma$. 
 Arrows,  showing 95\% confidence level upper limits,  are drawing when statistical significance of 
 a data point is lower than $1\sigma$. 
 The horizontal and vertical axes
 show the photon energy in MeV and count per unit energy interval, respectively. Detector
 responses have not been removed.
 }
 \label{fig:bg-sub_spectra_071213}
 \end{figure}

\begin{figure}
\noindent\includegraphics[scale=0.5]{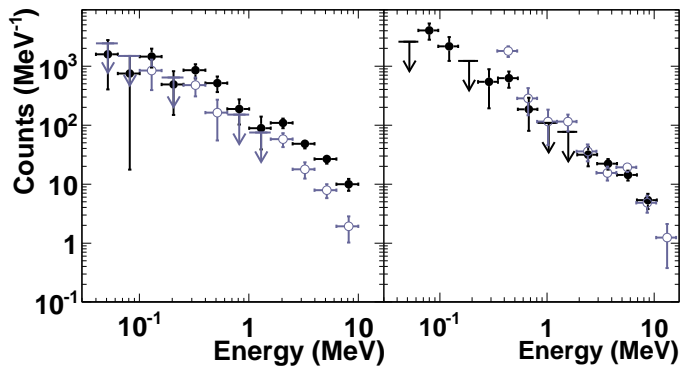}
 \caption{%
  The same as Fig.~\ref{fig:bg-sub_spectra_071213}, but for the 081225 event.
 }
 \label{fig:bg-sub_spectra_081225}
 \end{figure}

Figure~\ref{fig:bg-sub_spectra_071213} and Figure~\ref{fig:bg-sub_spectra_081225} show 
background-subtracted GROWTH spectra, obtained in the burst periods of 071213 and 081225, respectively. 
In either case, we accumulated the data over the burst period, and subtracted 
background spectra which were averaged over 10 minutes before and after the burst, 
although thunderstorms were ongoing during these time periods. 
This is to remove $<3$ MeV line-$\gamma$ rays induced mainly by radon decays, 
which increase the background level by up to twice.  
To examine how the background selection affects the final 
spectra (Figure~\ref{fig:bg-sub_spectra_071213} and Figure~\ref{fig:bg-sub_spectra_081225}), 
we subtracted an alternative background spectrum averaged over 5 minutes before and after the burst. 
However,  the background-subtracted spectra did not change by more than 
$\pm$10\% at $<1$ MeV, or $\pm$5\% at $ >1$ MeV. These are almost negligible compared 
with the statistical errors.

In both events, the background-subtracted spectra of Detector-A and Detector-B exhibit very 
hard continuum spectra, which clearly extend to 10 MeV.  
As shown in another GROWTH event 070106 reported previously~\citep{growth_2007}, and in
high-mountain observations~\citep{MtFuji_2009,norikura_2009, Chili_2010},  similar prolonged 
$\gamma$-ray emissions,  extending to 10 MeV or higher, were observed, and have been 
thought to be produced via bremsstrahlung. Thus, the present high-energy $\gamma$ rays must
also be produced via bremsstrahlung by electrons accelerated beyond 10 MeV. Given these
results, the present two events, together with the previous ones,  may be understood as 
manifestations of a common type of high-energy activity in thunderstorms.

As easily seen in Fig.~\ref{fig:bg-sub_spectra_071213} and Fig.~\ref{fig:bg-sub_spectra_081225}, 
the obtained spectra, in particular those of Detector-B, flatten in 0.8 $-$ 3 MeV,
even though they are not corrected for the detector responses. 
One of the causes of this flat is the Compton scattering: 
since the Compton scattering cross section in the atmosphere
increases as photon energy decreases toward 0.1 MeV, 
photons at low energies would experience stronger Compton degradation 
than higher-energy ones.
\subsection{Model fits}
Supposing that the burst $\gamma$ rays were produced in a source located at a certain distance
and propagated through atmosphere to reach our detectors, we may deduce the initial photon spectrum at
the source, and estimate the source distance, from the background-subtracted spectra. Since the Detector-A spectra are 
complicated due to the passive and active shielding effects by the BGO well, below
we analyze the Detector-B spectra.
According to numerical calculations~\citep{Roussel_1994,R_G_1996,Le_1999,Babich_MC2007},  an energy distribution function
of runaway electrons, generated under the RREA mechanism,
is expressed by a power-law function,  or more precisely, an exponentially cut-off 
power-law. Consequently, we assume an initial photon number spectrum as
\begin{equation}
f(\epsilon_\mathrm{p}) = \alpha \epsilon_\mathrm{p}^{-\beta}\exp{ (-\epsilon_\mathrm{p}/\epsilon_\mathrm{c})} \,\, (\mathrm{MeV^{-1}sr^{-1}}). \label{eq:epl}
\end{equation}
Here, $\alpha$ and $\beta$ 
are a normalization factor and a photon index, respectively, while $\epsilon_\mathrm{p}$ and $\epsilon_\mathrm{c}$ describe
the emitted photon energy and a cut-off energy in MeV, respectively. While this equation represents
an exponentially cut-off power law, it can also express a pure power-law by requiring $\epsilon_\mathrm{c} \rightarrow \infty $.
 
Below, let us estimate the source distance $d$ from our Detector-B data, 
as well as $\alpha$, $\beta$, and $\epsilon_\mathrm{c}$. 
In order to simulate the photon propagation in the atmosphere, 
we utilize EGS4~\citep{EGS4} embedded in CORSIKA 6.500~\citep{CORSIKA}.
In the CORSIKA simulation, the atmosphere consists of 
$\mathrm{N}_{2}$, $\mathrm{O}_2$, and $\mathrm{Ar}$ with the mole ratios of 
78.1\%, 21.0\%, and 0.9\%, respectively. The density of the atmosphere,
divided into 5 layers, depends exponentially on the altitude $h$, with a form of
$A + B  \exp{(-h/C)}$, with $A,B,C$ being model parameters. 
For example, at $h<4$ km, the model is specified as $A=-186.6\,\, \mathrm{g\, cm^{-2}}$,  
$B=1222.7\,\, \mathrm{g\, cm^{-2}}$, and $C=9.94$ km~\citep{CORSIKA_usermanual}. 
In addition, EGS4 can adequately treat electromagnetic processes in the relevant energy range of 
a few tens of keV to a few tens of MeV.

Mono-energetic photon simulations were
carried out for 33 incident energies from 50 keV to 100 MeV. 
The energy interval is set to 10 keV for 50 keV $-$ 90 keV, 100 keV for 100 keV $-$ 1 MeV,
1 MeV for 1 MeV $-$ 10 MeV, and 10 MeV for 10 MeV $-$ 100 MeV. For one
mono-energetic photon simulation, one million photons were vertically injected to the atmosphere
from a fixed source distance. In reality, 20 source distances from 20 m to 2000 m were 
applied for one mono-energetic simulation. Then, we saved the energy, angle, and 
species of all of photons and particles that arrive at the observatory level (40 m above sea level).

\begin{figure}
\noindent\includegraphics[scale=0.25]{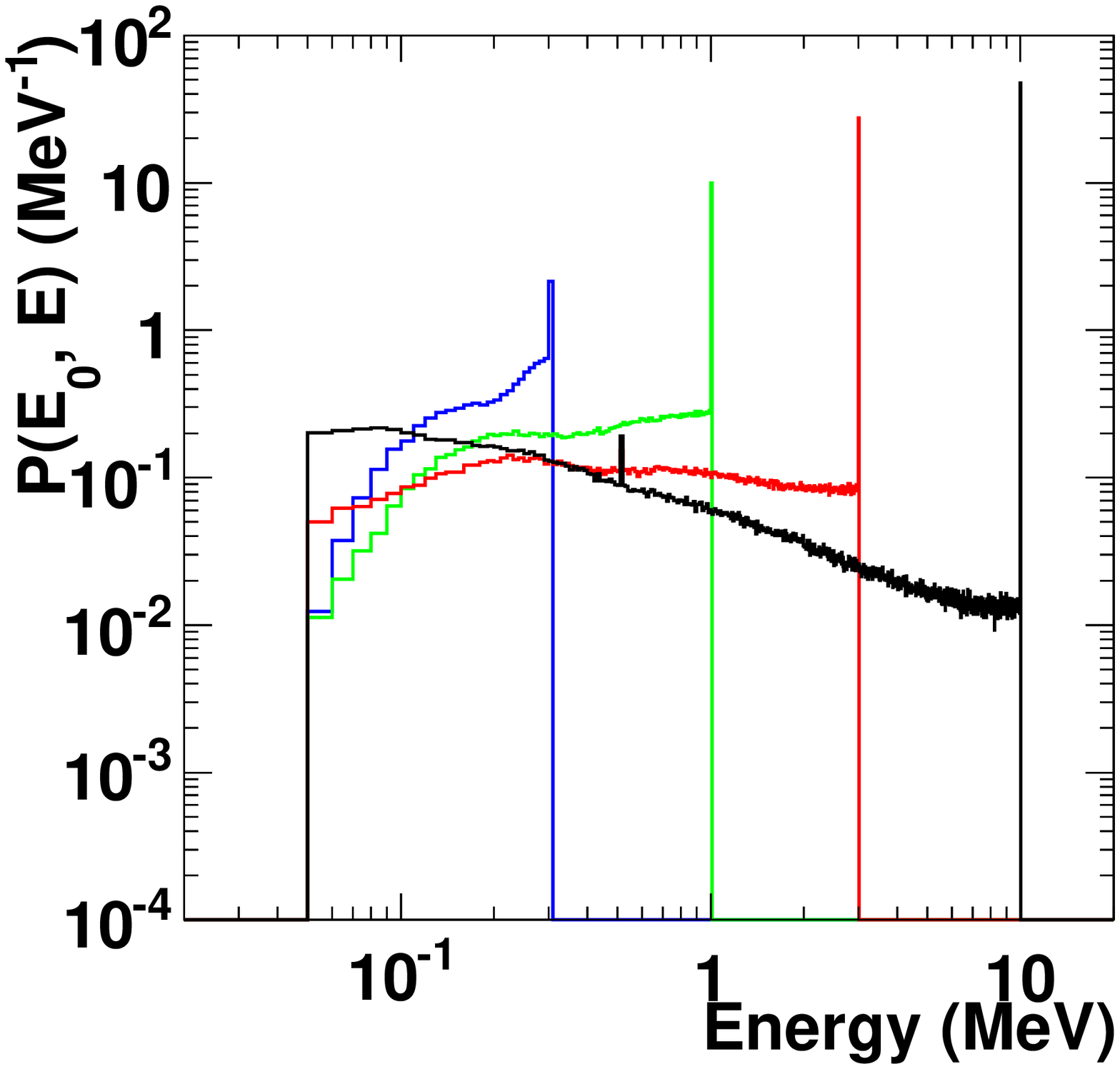}
\noindent\includegraphics[scale=0.25]{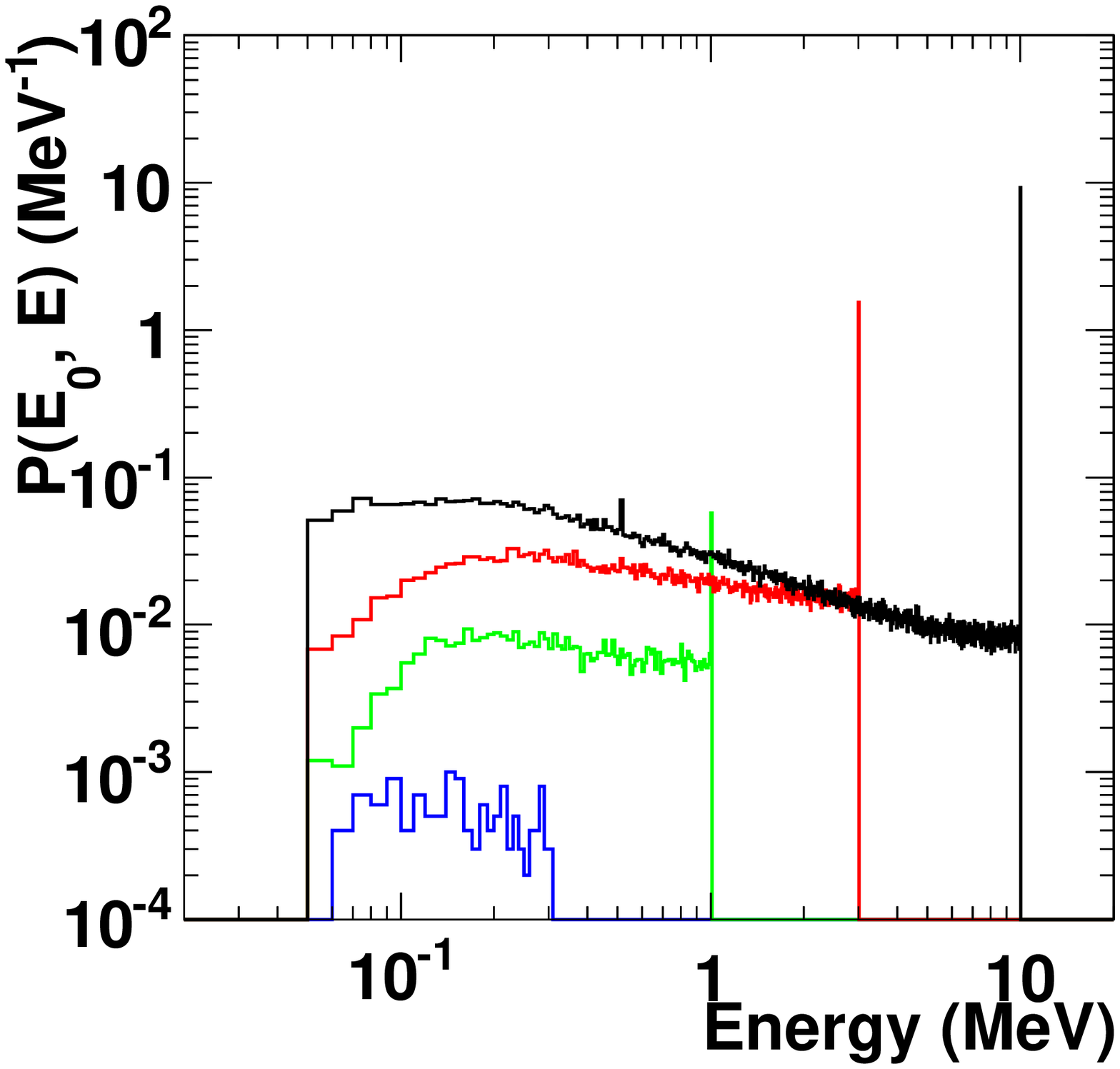}
\noindent\includegraphics[scale=0.25]{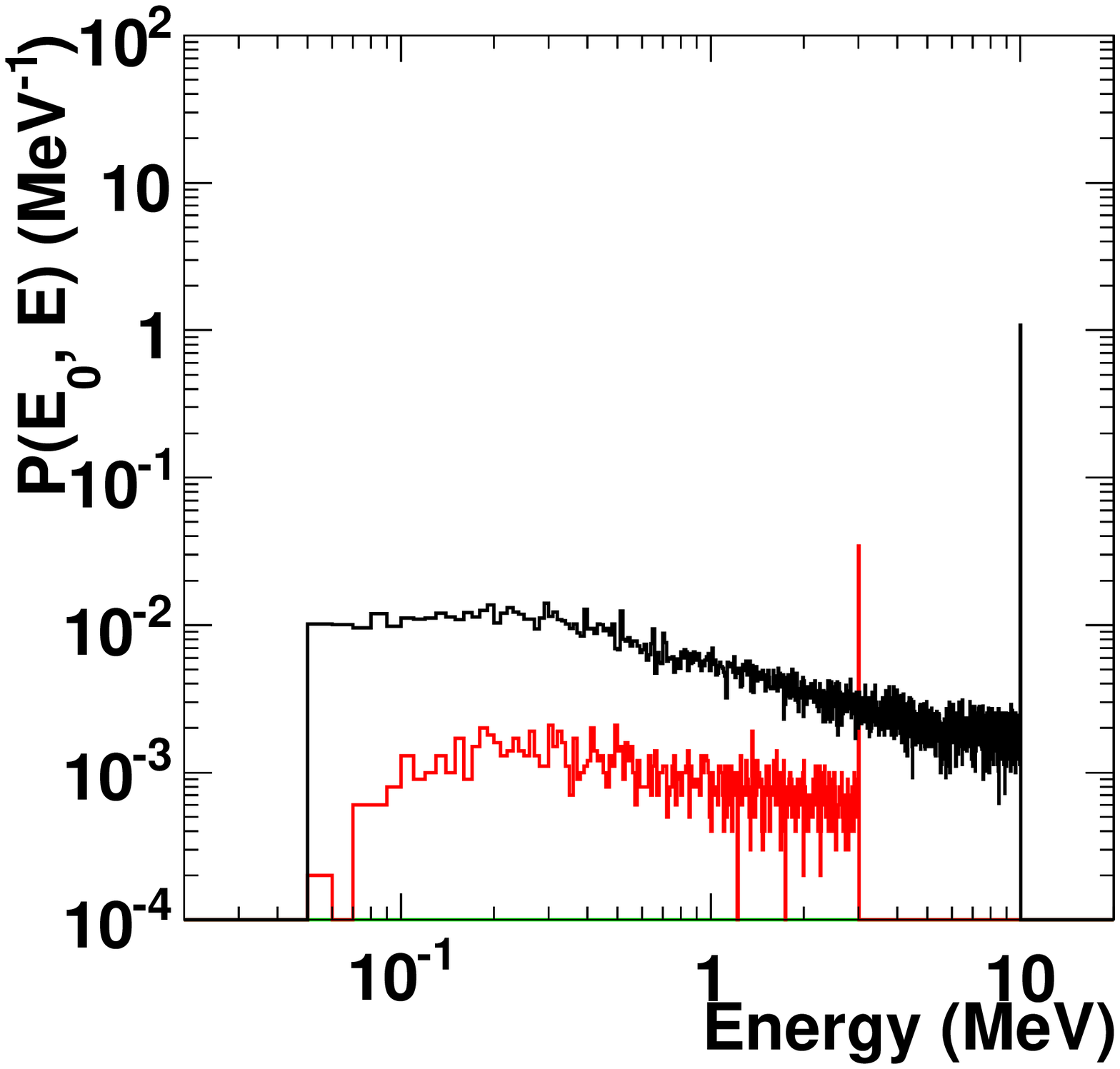}
 \caption{%
Photon spectra at the observatory derived from Monte Carlo simulations. 
Three panels correspond to $d=300$ m (left), 1000 m (middle), and 2000 m (right). 
Different colors  denote incident photon energies, $E_0=$ 0.3 (blue), 1 (green), 3 (red), and 10 MeV (black)
in all panels. Abscissa shows
the photon energy at the observatory, while ordinate represents probability density function.
 }
 \label{fig:atte_spe}
 \end{figure}

Figure~\ref{fig:atte_spe} indicates three representative sets of simulated photon spectra, 
propagating over $d=$ 300 m ($36\, \mathrm{g\, cm^{-2}}$),  1000 m ($120\, \mathrm{g\, cm^{-2}}$), 
and 2000 m ($220 \, \mathrm{g\, cm^{-2}}$),
with the numbers in parentheses giving air mass calculated by the above exponential formula.
Punch-through photons, which suffer no interactions 
with air molecules, appear as a strong peak at the highest end of each photon spectrum, while scattered ones
form a continuum toward lower energies. As the distance increases, 
the punch-through photons and the scattered continuum are both strongly attenuated, 
in particular toward lower energies, due primarily to Compton scattering.
For instance,  the survival probability for 10 (1) MeV punch-through
photons to propagate over $d=2000$ m is only 0.02 ($10^{-5}$) times that over $d=300$ m.
Note that as discussed later in Sec.~\ref{dis:enum}, a long-duration burst probably changes in the burst period 
its viewing angle relative to a beam axis of electrons accelerated in thunderclouds. Thus, 
the calculated photon spectra here would vary according to the changes, and hence
they will be treated in this work as ones averaged over different viewing angles.

Convolving the simulated photon spectra with the 
detector responses, we can obtain a model-predicted spectrum to be observed
by the NaI and CsI scintillators.
Finally, we convolve these model predictions with the assumed source photon spectrum,
eq.(\ref{eq:epl}), and fit the predictions simultaneously to
the background-subtracted NaI and CsI spectra (right panels of Fig.~\ref{fig:bg-sub_spectra_071213} and 
Fig.~\ref{fig:bg-sub_spectra_081225}). 
Then, the model parameters, such as $\alpha$, $\beta$ and $\epsilon_\mathrm{c}$,   
can be determined so as to minimize the fit $\chi^2$.  

\begin{figure}
\noindent\includegraphics[scale=0.50]{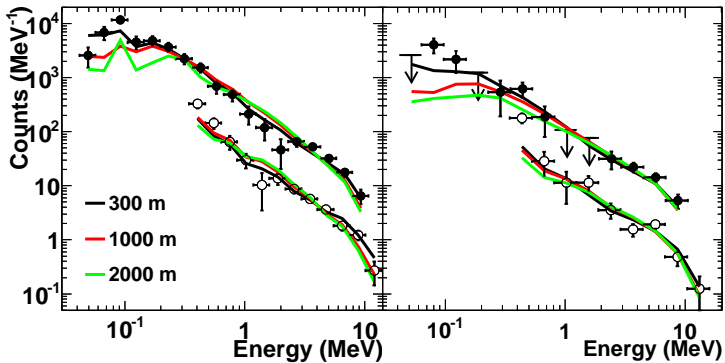}
 \caption{%
 The photon spectra observed by the NaI (filled circles) and CsI (open circles) 
 scintillators of Detector-B,  compared with calculations for assumed source distances of 
 300 m (black), 1000 m (red), and 2000 m (green). Left panel shows 071213, while right one indicates
 081225.  For clarity, the CsI data and the corresponding model  spectra are 
 multiplied by 0.1. The horizontal and vertical axes  show the photon energy in 
 MeV and counts in each bin,  respectively.}
  \label{fig:bg_sub_with_models}
 \end{figure}

\begin{figure}
\noindent\includegraphics[scale=0.50]{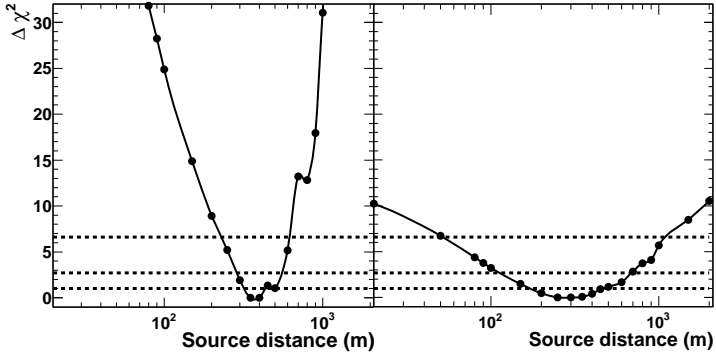}
 \caption{%
 The values of  $\Delta \chi^2=\chi^2 - \chi^2_\mathrm{min}$, plotted as a 
 function of the assumed source distances (black circles).   Left and right panels show 071213 and
 081225, respectively. Black curves show
 smoothing lines. Horizontal dashed lines from bottom to top correspond to
 68\%, 90\%, and 99\% confidence level.}
  \label{fig:chi2map}
 \end{figure}

Shown in Figure~\ref{fig:bg_sub_with_models} are three representative model fits to the 
spectra of 071213 and 081225, assuming a power-law model.  The choice of $d$ of
300, 1000, and 2000 m in Fig.~\ref{fig:bg_sub_with_models}, respectively gave $\chi^2$ values as
$48.4$, $77.5$, and $116$ for 071213, and $40.9$, $46.7$, and $51.4$ for 081225.
By changing $d$ and repeating the fitting, we obtained $\chi^2$ curves as shown in
Figure~\ref{fig:chi2map}, together with the $\chi^2$ minima as 46.5 at $d=400$ m for 071213,
and $40.9$ at $d=300$ m for 081225.  
From these figure, we can constrain source distance.  Table~\ref{tab:spepara_071213} and 
Table~\ref{tab:spepara_081225} summarize the best-fit parameters for 071213 and 081225, respectively, 
together with the constrained source distance. 
Also, low-energy parts ($<300$ keV) of
the two spectra are found to play an important role to determine $d$. 
If we use the background averaged over the 5 min intervals
(instead of 10 min), the source distances become $350 $ m for 071213, and $300$ m for 081225.
Thus, the distance is not affected significantly by the systematic background uncertainty.

The NaI and CsI spectra have been explained, in either event, by a common set of model
parameters, although the fits are not necessarily good enough. 
 The cut-off energy $E_\mathrm{c}$ was constrained to be rather high
with relatively large errors. Thus, our data do not provide evidence for
spectral cut-off in either event.
In agreement with this, the two spectral models, a power law and an exponentially cut-off power law, 
gave similar goodness of fits in both events. Importantly, the source distance 
have been constrained with a reasonable accuracy.

As another attempt,  we tentatively fixed $E_\mathrm{c}$ at 7 MeV, which is the expected average 
kinetic energy of runaway electrons (not of bremsstrahlung photons),  and repeated the model
fitting. Then, the fit became worse in both events (4th column of Table~\ref{tab:spepara_071213} and 
Table~\ref{tab:spepara_081225}). Therefore, the initial photon spectrum is again inferred to extend 
beyond $\sim$ 7 MeV.  
\section{Discussion}\label{dis}
\subsection{Source heights}\label{dis:height}
Assuming a power-law function at the source, the $\gamma$-ray spectra of 071213 
and 081225 suggest that the sources are located at 290 $-$ 560 m ($35 -67\,\, \mathrm{g\, cm^{-2}}$) 
and 120 $-$ 690 m ($14 - 82 \,\, \mathrm{g\, cm^{-2}}$) above our system, respectively, both at 90\% confidence 
level. In fact, these constraints are in good agreement with the known
heights of winter thunderclouds in this area. 
Winter thunderclouds and winter lighting observed at the coastal area
of Japan Sea exhibit many features that have hardly been found from those in summer seasons
and/or in other areas (e.g. \citet{RU_lightningbook} and references therein). These include rather 
low altitudes of the development of these thunderclouds.
Actually, \citet{GN_isolate_1992} conducted video observations of winter lightning
at the same Niigata Prefecture as our experimental site, 
and reported that the visible bases of winter thunderclouds 
are typically located at 200 m $-$ 800 m above sea level.  Also, a recent
numerical calculation done by \citet{Babich_MC2010} shows that
another GROWTH event~\citep{growth_2007} may be produced 
at a source height of 0.5 $-$ 2 km, and hence generally agrees with the present results.

These height estimations provide an additional clue to the possible electron contributions
to the detected plastic signals (Sec. \ref{res:comp}). As argued so far, electrons are considered 
to be accelerated in these thunderclouds to at least 10 MeV, probably a few tens of MeV. 
Since such electrons have a range of $< 100$ m at near the sea level,  they would hardly reach
our system, even if a range straggling is taken into account. Therefore, 
it is reasonable that the electron flux incident on our system, if any, 
was much lower than that of photons.
 
Unlike the present sea-level observations, some high-mountain experiments,  conducted
at Mt. Norikura  (2770 m) in Japan~\citep{norikura_2009} and Mt. Aragatz (3250 m) in
Armenia~\citep{Chili_2010}, have detected primary electrons in long duration 
events (numbers in parentheses  indicate altitudes of observatories). 
\citet{norikura_2009} estimated the source height as 60 $-$ 130 m (90\% confidence level),
while \citet{Chili_2010} evaluated it as 100 $-$ 150 m. 
These low source heights, which are comparable to or shorter 
than the expected electron range, can naturally explain their electron detections.
\subsection{Extent and motion of the $\gamma$-ray beams}\label{dis:extent}
Measuring electric-field structure of winter thunderclouds, 
\citet{KM_1994} revealed that tripole electrical structures, 
which consist of positive, negative and positive layers from top to bottom, appear 
at mature stages of winter thunderclouds. Then, they observed 
the tripole structures to last for $<10$ minutes in early 
or late winter, while less than several minutes in midwinter. Since the present two 
events were observed in midwinter,  
the measured burst periods of 84 sec of 071213 and 72 sec of 081225 are 
consistent with their observations, if the burst durations represent the lifetime
of electric fields.

\begin{figure}
\noindent\includegraphics[scale=0.6]{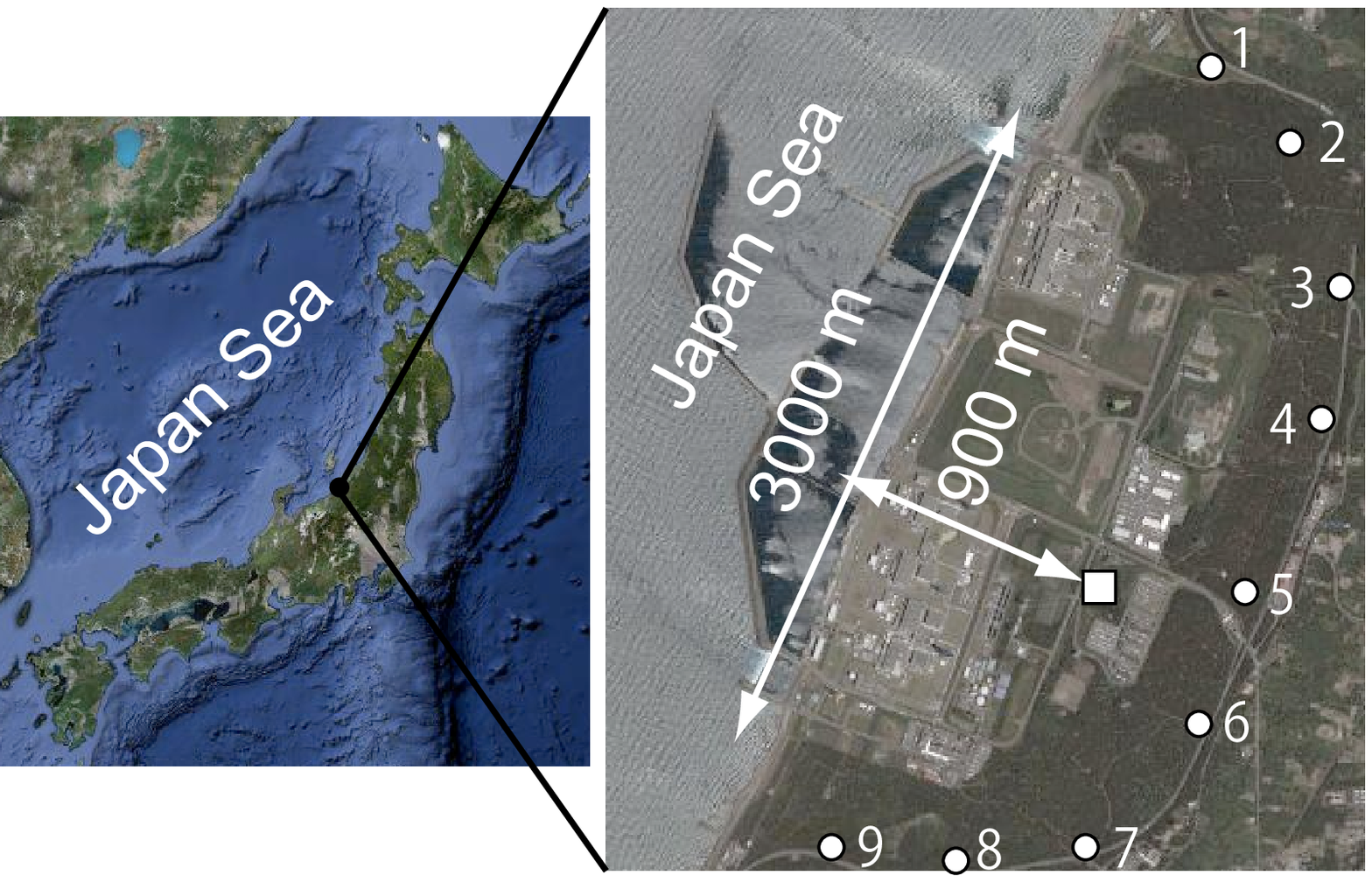}
 \caption{The location of the Kashiwazaki-Kariwa nuclear power plant (left) and its bird view (right).
 A filled square in the right panel represents the GROWTH experimental site, while 9 filled circles show locations
 of radiation monitors.  Each original image is taken from Google Map.}
 \label{fig:location}
 \end{figure}

\begin{figure}
\noindent\includegraphics[scale=0.5]{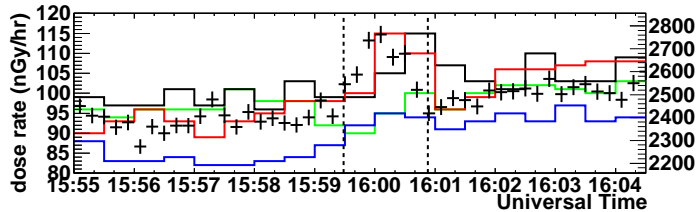}
 \caption{%
Radiation dose rates per 30 sec (left ordinate) obtained by ion chambers of radiation monitors over
15:55 $-$ 16:05 UT on  2007 December 13.  
Different colors specify the radiation monitors numbered as 4 (green), 5 (black), 6 (red), and 7 (blue) in Fig.~\ref{fig:location}. 
Superposed on them, crosses show the $>$ 40 keV count history per 12 sec of the NaI detector of Detector-B (right ordinate). 
Vertical dashed lines represent the defined start and end time of the 071213 event.
 }%
 \label{fig:mp_vs_nai_detb}
 \end{figure}

Figure~\ref{fig:mp_vs_nai_detb} shows dose variations on 2007 December 13, measured  by 
the nearest and the second nearest radiation monitors of the power plant (5 and 6 in Fig.~\ref{fig:location}, 
black and red lines in Fig.~\ref{fig:mp_vs_nai_detb}, respectively). 
The two monitors gave moderate dose increases for $\sim1$ minute or less around the GROWTH event. 
By examining the GROWTH data of 071213 burst (crosses in Fig.~\ref{fig:mp_vs_nai_detb}), 
as well as dose rates of the second nearest monitor (red line) obtained for 15:59:30 $-$ 16:01:00 UT
and that of the nearest one (black line) obtained over 16:00:00 $-$ 16:01:30 UT,
peak times of their enhancement can be evaluated as 15:59:48 ($\pm6$ sec) UT, 15:59:58 ($\pm15$ sec) UT, 
and 16:00:27 ($\pm15$ sec) UT.
Thus, referring to the GROWTH data, 
the second nearest monitor increased in its dose rates with a small delay of $10\pm16$ sec (or
almost simultaneously), while the nearest one with a larger delay by $39 \pm 16$ sec.
The two monitors are located at a distance of 500 $-$ 600 m from the GROWTH system. For reference, 
data of the other two radiation monitors (4 and 7 in Fig.~\ref{fig:location}) exhibited no apparent increases 
(green and blue lines in Fig.~\ref{fig:mp_vs_nai_detb}).
As for 081225, data of those radiation monitors were unavailable due to some data-storage problem. 

These simultaneous and delayed detections by the two radiation monitors have two important implications.
One is that the $\gamma$-ray emission from thunderclouds is likely to have illuminated a rather limited 
area, spreading over $\sim600$ m on the ground.
This kind of effect was also suggested by five radiation monitors (1$-$5 in Fig.~\ref{fig:location}),
on the occasion of the other GROWTH event~\citep{growth_2007}, and another experiment
conducted on the same coastal area~\citep{MONJU_2002}.
The other is that the $\gamma$-ray emitting region moved, 
presumably together with the thunderclouds. 

From data of Japan Meteorological Agency, it is found that south-west wind was blowing during 
10 minutes including the burst period of 071213. 
Thus, the south-west direction can naturally explain the delay of the nearest monitor, if the
 $\gamma$-ray emitting region moved together with the thunderclouds. 
Then, the wind velocity was on average $360\, \mathrm{m\, min^{-1}}$, with the maximum of $720 \, \mathrm{m\, min^{-1}}$.
Projecting, to south-west axis, the distance between the GROWTH system and the nearest monitor, $\sim500$ m, 
and dividing the projected distance, $\sim350$ m, by its delay, $39\pm16$ sec,  we obtain an average moving velocity of 
the emitting region as $540 \pm 220\,\, \mathrm{m\, min^{-1}}$.
Thus, the estimated moving velocity is generally consistent with the wind velocity.

Given above discussions, we may assume that the winter thunderclouds moved 
from the Japan sea in south-west side to the inland in north-east side.
Then, a short-lived tripole structure appeared in a thundercloud, and accelerated 
ambient fast electrons toward the bottom positive layer. The accelerated electrons
emitted $\gamma$ rays toward the ground, which 
the GROWTH system and the two radiation monitors detected when
the beam passed over them. The differences in the statistical significance of detections
between the GROWTH system and the radiation monitors may be due to 
different positions and effective viewing angles relative to the accelerated electron beam axis in the thundercloud, 
and to the differences in their sensitivity.
The 081225 event is considered to have occurred under similar conditions, because west winds, 
with almost the same velocity as in the 071213 case, were blowing at that time.
However, 
it is presently unclear whether the $\gamma$-ray emission ceased when 
the tripole structure disappeared, or when the $\gamma$-ray beam moved 
away from the GRWOTH system as the thunderclouds moved.
\subsection{The number of relativistic electrons in thunderclouds}\label{dis:enum}%
Using the initial photon energy spectrum $f(\epsilon_\mathrm{p})$ of Eq. (1) as quantified in Table~\ref{tab:spepara_071213} and Table~\ref{tab:spepara_081225}, 
we can estimate the number of relativistic electrons radiating the observed $1-10\, \mathrm{MeV}$ $\gamma$ rays
via bremsstrahlung, as
\begin{equation}
N_\mathrm{e} \sim \frac{2\pi}{H}\int_{1}^{10} \! dK_\mathrm{e} \int_{1} ^{K_\mathrm{e}} \! d\epsilon_\mathrm{p} \int_{0}^{\theta_\mathrm{max}}\frac{f(\epsilon_\mathrm{p})}{\eta(K_\mathrm{e}, \epsilon_\mathrm{p},\theta)}\sin{\theta}d\theta.
\label{eq:Ne}
\end{equation}
Here, $\eta(K_\mathrm{e}, \epsilon_\mathrm{p},\theta)$ is the probability per 1 $\mathrm{g\,cm^{-2}}$ with which 
an electron with a kinetic energy $K_\mathrm{e}$ emits a bremsstrahlung photon with an energy $\epsilon_\mathrm{p}$ 
at an angle $\theta$ with respect to the electron-beam axis~\citep{KM_1959}, and 
$H$ denotes the vertical length of the acceleration region. Since this $H$ is unavailable from the
present observations like in \citet{norikura_2009}, we assume either 
$H=$  300 m or 1000 m, corresponding to $35\, \mathrm{g\, cm^{-2}}$ and $110\,\mathrm{g\, cm^{-2}}$, 
respectively. These assumptions are based on intracloud observations of
X rays using a ballon-born detector, which showed that a high electric field region, to produce a significant flux in $3-120$ keV energy range, 
has a vertical extent of $\sim$ 500 m, at altitudes of 3.7 $-$ 4.2 km~\citep{Eack_1996}. 

We further assumed that the electric-filed strength in the acceleration 
region is 300 $\mathrm{kV\, m^{-1}}$, which 
is slightly higher than the threshold (at 1 atm) to cause the runaway electron avalanches. When 1 MeV electrons 
are accelerated from the top of this acceleration region to the bottom, they
will gain energies of 15 MeV for $H=300$ m, and 19 MeV for $H=1000$ m.
Therefore, the assumed electric-field strength, together with the assumed vertical length,  
is sufficient to produce 10 MeV photons via bremsstrahlung. 

To calculate eg.(\ref{eq:Ne}), we further need to specify $\theta$; this is suggested
to be relatively small, from the obtained photon spectra.
In practice, the angle of prolonged $\gamma$-ray event may vary according to the motion 
of thunderclouds. Thus, we adopt $15^{\circ}$ or $30^{\circ}$ as $\theta_\mathrm{max}$. 
As listed in Table~\ref{tab:Ne},  these assumptions, together with eq. (\ref{eq:epl}), give $N_\mathrm{e} = 10^{9}-10^{11}$. 
Similar estimations for other long-duration $\gamma$-ray bursts have given $N_\mathrm{e}=10^8-10^{12}$~\citep{growth_2007,norikura_2009, Chili_2010}.
Thus,  long-duration $\gamma$-ray bursts appear to be emitted by a similar number of relativistic electrons.
\subsection{Relation between the bursts and the RREA mechanism}\label{dis:CR}%
A possible source of energetic seed electrons to cause the RREA
can be attributable to secondary cosmic rays~\citep{RREA_1992}. The cosmic-ray flux above 1 MeV, at the 
presently relevant altitudes of $<$1 km, is $I_{0} \sim 200\, \mathrm{m^{-2}s^{-1}}$~\citep{CR_text}. 
Considering the measured burst periods of $70-80$ sec and the 30 $-$ 40 sec delay of one radiation monitor from the 071213 event, 
we presume that the acceleration region has a horizontal length of $L\sim600$ m at most, as
judged from the extent of the $\gamma$-ray beam of 071213.
Possibly, an actual extent of the acceleration region in thunderclouds would be shorter than this 600 m, because 
the $\gamma$-ray beam would diverge due to multiple scatterings of the emitting electrons and Compton
scatterings of the emitted $\gamma$ rays. We may also consider that an acceleration
region is sustained in thunderclouds at least for 100 sec.
Accordingly, the number of cosmic rays $S_{0}$,  entering the acceleration region,
is described as
\begin{equation}
S_{0} = 7.2\times10^9 \times (L/600\,\,\mathrm{m})^2\times \Delta t/100\,\, \mathrm{sec}.
\end{equation}

Based on the RREA mechanism, 
the total number of relativistic electrons at the end of an acceleration 
region, $N_\mathrm{RREA}$, is estimated as 
\begin{equation}
N_\mathrm{RREA} = S_{0}  \exp(\delta), \, \, \delta =  \int_{0}^{H}\frac{dz}{\lambda}.
\end{equation}
The length parameter $\lambda$ is given as
\begin{equation}
\lambda = \frac{7300 \, \mathrm{kV}}{E - (276\, \mathrm{kV\, m^{-1}})n} \,\, \mathrm{m},
\end{equation}
where $E$ is the electric-field strength in $\mathrm{kV\, m^{-1}}$ and $n$ denotes
the air density relative to that at 1 atm. This formula is valid for
$300 - 3000\,\, \mathrm{kV\, m^{-1}}$~\citep{Dwyer_MC2003}.
Assuming $E = 300 \,\,\mathrm{kV\, m^{-1}}$ gives $\lambda \sim 300 \, \mathrm{m}$ at $P=1 \,\,\mathrm{atm}$.
In practice, $E$ may be somewhat lower than 300 $\mathrm{kV\, m^{-1}}$, because
$P$ during thunderstorms would be usually lower than 1 atm due to lower pressure system, and hence
gives $n<1$. Since a uniform field, $\delta = H/\lambda$, gives $N_\mathrm{RREA} = S_{0}\exp{(H/\lambda)}$,  
the factor $\eta = \exp{(H/\lambda)}$ is regarded as the avalanche multiplication factor, and becomes 3 and 30 
for $H$ = 300 m and 1000 m, respectively. 
As a result,  we obtain
\begin{equation}
N_\mathrm{RREA} = 2.2\times10^{11}\times(L/600\,\,\mathrm{m})^2\times \Delta t/100\,\, \mathrm{sec} \times \eta/30.
\label{eq:Nrrea}
\end{equation}
We thus obtain $N_\mathrm{RREA} = 10^{10} -10^{11}$,  which agrees generally with the derived $N_\mathrm{e}=10^{9}-10^{11}$.
Thus, the standard RREA process can explain at least the present two prolonged bursts.

In the above estimation, we assumed an electric field is slightly higher than
the RREA threshold. However, a weaker field below this threshold might suffice to 
produce prolonged $\gamma$-ray emission.
In reality, a 30 $-$ 120 keV x-ray flux continuously increased while an electric field is lower by $30\% - 60\%$
than the RREA threshold~\citep{Eack_1996}.
This quasi-static moderate-level field might be accomplished by 
e.g. a charging mechanism of thunderclouds. 
\subsection{Comparisons with TGFs}\label{dis:spectra}%
The derived $N_\mathrm{e}$ (sec.\ref{dis:CR}) is more than five orders of magnitude lower than the number of relativistic
electrons expected from TGF observations, e.g. $10^{16}-10^{17}$~\citep{DS_TGF_2005}.  
This huge number of relativistic electrons in TGFs may be generated by relativistic feedback mechanism, 
involving positrons and X rays propagating in the opposite direction to runaway electrons~\citep{Dwyer_MC2007,Dwyer_TGF2008}.
Since the estimated $N_\mathrm{e}$ of the present bursts is in generally agreement with $N_\mathrm{RREA}$ expected from the
simple RREA mechanism, we conclude that at least the present two events do not require an intense feedback process.

\begin{figure}
\noindent\includegraphics[scale=0.4]{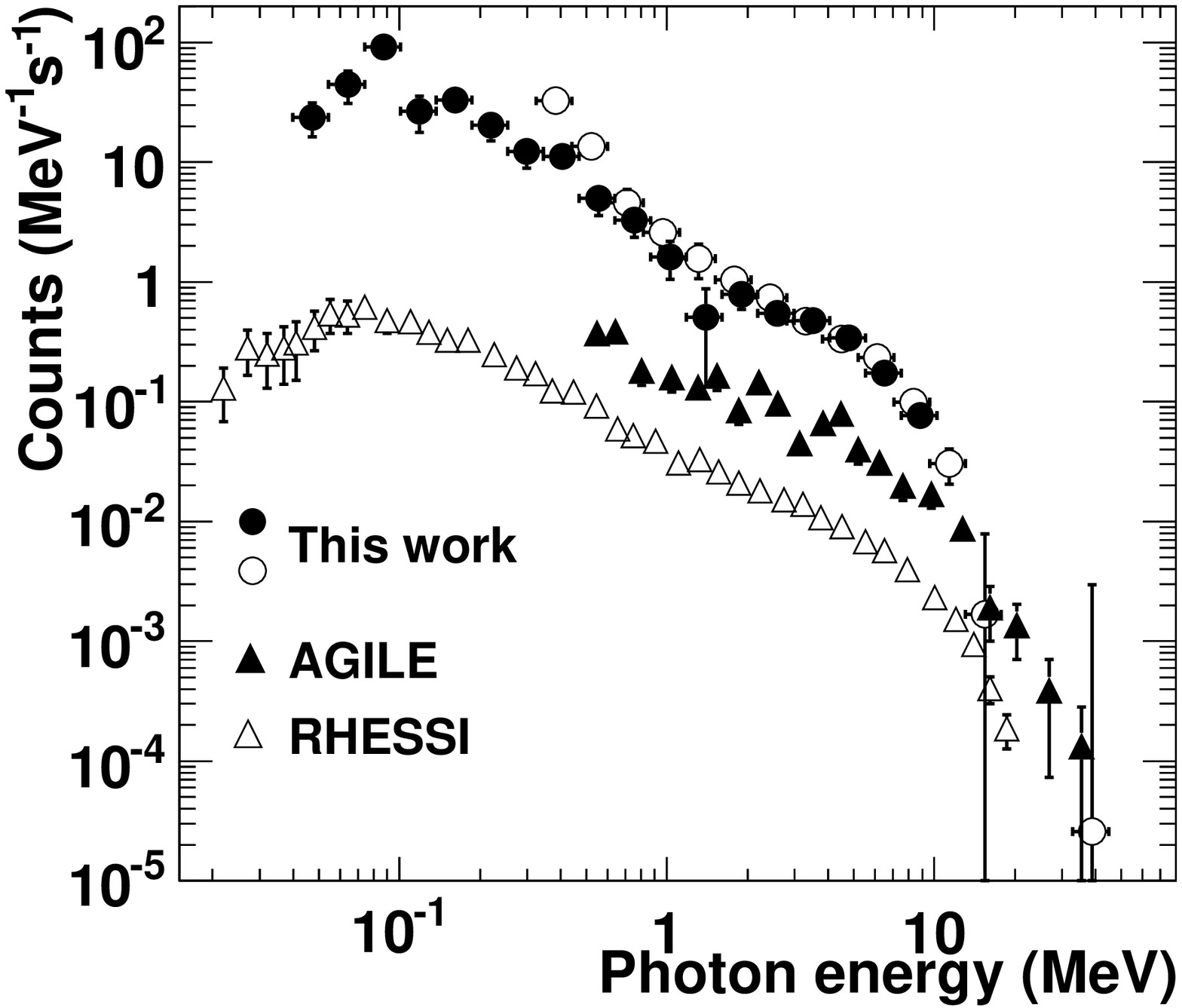}
 \caption{%
Cumulative spectra of the three GROWTH events from Detector-B [NaI and CsI (filled and open circles, respectively)], 
compared with summed TGF spectra by  RHESSI and AGILE.
The latter two spectra are adopted from Fig.2 of \citet{DS_TGF_2005} and Fig.5 of \citet{AGILE_MCAL}.
For clarity, the RHESSI and AGILE spectra are multiplied by $1\times10^{-5}$ and $ 1\times10^{-4}$, respectively.
The vertical and horizontal axes represent counts in $\mathrm{MeV^{-1}s^{-1}}$
and photon energy in MeV, respectively. Errors assigned to the GROWTH and AGILE data
are statistical ones, while those for RHESSI include systematic uncertainty of background estimation~\citep{DS_TGF_2005}.
 }%
 \label{fig:growth_2007_rhessi}
 \end{figure}

In order to better characterize $\gamma$-ray spectra of long-duration events,  we stacked count spectra 
over three bursts, namely, the present two ones and 070106~\citep{growth_2007}. Figure~\ref{fig:growth_2007_rhessi} 
compares the summed GROWTH spectrum with averaged TGF ones obtained by two independent satellites;
one sums  289 events measured by Reuven Ramaty High Energy Solar Spectroscopic Imager (RHESSI)~\citep{DS_TGF_2005},
while the other averages over 34 events observed by the Astrorivelatore Gamma a Immagini Leggero (AGILE) satellite~\citep{AGILE_MCAL}.
Thus, the composite GROWTH spectrum is rather similar in shape to those from TGFs,
although three spectra all include detector responses.
This resemblance is consistent with our basic standpoint~\citep{growth_2007,norikura_2009}
that the long-duration $\gamma$ rays are emitted by the same bremsstrahlung process as 
TGFs. 

On a close comparison, the TGF spectra, especially the AGILE one,  
appear to have a higher cut-off energy than the GROWTH spectrum. 
This may be attributable to a difference in the electric potential operating 
in an acceleration region~\citep{DS_TGF_2005}. 
Thus, electrons accelerated in a much lower atmospheric density at the production sites of TGFs, 
$15-40$ km~\citep{DS_TGF_2005,Carlson_TGF_cal,BATSE_TGF_2008}, would propagate 
through a longer distance, which gives a higher electric potential, and gain higher energies because of
a smaller ionization loss per unit length. 

\section{Summary}\label{conc}
The GROWTH experiment observed two long-duration $\gamma$-ray emissions from winter thunderstorms on 
2007 December 13 and 2008 December 25.
The photon spectra obtained in both events clearly extends to 10 MeV, 
and are consistent with a scenario that accelerated electrons produce, via  bremsstrahlung, 
the observed  $\gamma$ rays. 
Adopting a power-law function as the initial photon spectrum at the source, we have constrained 
the source distance as 290 $-$ 560 m for 071213 and 110 $-$ 690 m for 081225,
both at 90\% confidence level.
These constraints agree with visible-light observations, which show
that the bottom of winter thunderclouds is usually located at 200  $-$ 800 m above sea level~\citep{GN_isolate_1992}.
We have shown a possibility that the observed $\gamma$-ray beams move with winter thunderclouds,
and spread over $\sim 600$ m. 

We estimated the number of relativistic electrons to cause the present prolonged $\gamma$-ray emissions as 
$10^{9} - 10^{11}$. These are in general agreement with
those expected from the standard RREA mechanism triggered by secondary cosmic rays.
The cumulative GROWTH spectrum,  summed over the present two ones and another GROWTH event~\citep{growth_2007}, 
was found to be similar in basic spectral features with the averaged TGF spectra~\citep{DS_TGF_2005,AGILE_MCAL}. 

\begin{acknowledgments}
We thank members of radiation safety group at Kashiwazaki-Kariwa 
power station, Tokyo Electric Power Company, 
for supporting our experiment.  
This work is supported in part by the Special Research 
Project for Basic Science at RIKEN, titled ``Investigation of Spontaneously 
Evolving Systems'', and the Special Podoctoral Research Project for Basic Science at RIKEN. 
The work is also supported in part by Grant-in-Aid for Scientific Research (S), No.18104004, and
Grant-in-Aid for Young Scientists (B), No. 19740167. 
\end{acknowledgments}

%
%
%
%
%
%
%
%
%
%

%
%
\end{article}
\clearpage
%
%
%
%
%
%
%
\clearpage
\begin{table*}
 \caption{The count enhancements and the corresponding photon number flux of 071213 and 081225.}
 \label{tab:netcounts_flux}
\begin{tabular}{ccccccc}
{} &\multicolumn{3}{c}{071213} & \multicolumn{3}{c}{081225} \\ \hline
 $\Delta E$ (MeV)  &Det. A\tablenotemark{a}        & Det. B\tablenotemark{b}      & flux\tablenotemark{c} ($\mathrm{\times 10^{-2}cm^{-2}s^{-1}}$)    & Det.A\tablenotemark{a}     & Det. B\tablenotemark{b}   & flux\tablenotemark{c} ($\mathrm{\times 10^{-2}cm^{-2}s^{-1}}$) \\
$0.04 - 0.3$      & $530 \pm 70$          &  $1120 \pm 110$   & $20.2 \pm 1.9$& $300 \pm 60$&$390 \pm 90$    & $6.1 \pm 1.7$ \\
$0.3   - 3$         & $900 \pm 80$              &  $1660 \pm 140$   & $22.2 \pm 2.3$& $400 \pm 60$  &$920 \pm 110$  &  $7.6 \pm 2.1$\\
$3 - 10$            & $410 \pm 30$              &  $370 \pm 20$       & $10.5 \pm 1.0$& $ 180\pm 20$  &$178 \pm 16$    &  $6.8\pm 0.9$\\ \hline
\end{tabular}
\\
\tablenotetext{a}{Sum of the two NaI detectors.}
\tablenotetext{b}{The $0.04-0.3$ MeV count correspond to the NaI detector, while the others a sum of the NaI and CsI detectors.}
\tablenotetext{c}{The flux is calculated by the data of Detector-B.}
\end{table*}
%
%
%
%
\begin{table}
\caption{The obtained spectral parameters and $\chi^2$/{d.o.f.} of the 071213.}
 \label{tab:spepara_071213}
\begin{tabular}{lccc}
{}                                                               &P. L. \tablenotemark{a}           & Exp. P.L. \tablenotemark{b}    & Exp. P.L. (fix) \tablenotemark{c}   \\ \hline
 $\alpha \,\,(\mathrm{MeV^{-1}sr^{-1}})$  &  $(1.25\pm0.03)\times10^{11}$    & $(7.19\pm0.02)\times10^{10}$ & $(4.9\pm0.6)\times10^{9}$   \\
$\beta$                                                   &  $2.03\pm0.02$                        & $1.88 \pm 0.01$                   &  $1.2\pm0.1$  \\
$\epsilon_\mathrm{c}$ (MeV)                          &  $-$                                       &  $50\pm 20$                     & 7    \\
$\chi^2_\mathrm{min}/${\it d.o.f.}\tablenotemark{d}           & 46.5/28 (1.6\%)                  &   46.0/27 (1.3\%)                              &  49.6/28 (0.72\%)   \\
$d $ (m) \tablenotemark{e}                     &$400^{+160}_{-110}$ &$350^{+210} _{-240}$   & $150^{+60}_{-70}$ \\ \hline
\end{tabular}
\\
\tablenotetext{a}{Power law model.}
\tablenotetext{b}{Exponentially cut-off power law model.}
\tablenotetext{c}{Exponentially cut-off power law with $\epsilon_\mathrm{c}$ being fixed at 7 MeV.}
\tablenotetext{d}{Values in parentheses represent the probability with the given $\chi^2_\mathrm{min}$ and d.o.f..}
\tablenotetext{e}{Quoted errors of $d$ are 90\% confidence values, while other errors are 68\% ones. }
\end{table}
%
%
%
%
\begin{table}
 \caption{The obtained spectral parameters and $\chi^2$/{d.o.f.} of 081225.}
 \label{tab:spepara_081225}
\begin{tabular}{lccc}
   {}                                                            &P. L.            & Exp. P.L.      & Exp. P.L. (fix)   \\ \hline
$\alpha \,\,(\mathrm{MeV^{-1}sr^{-1}})$  &  $(1.21\pm0.09)\times10^{9}$      &$(6.4\pm1.0)\times10^{9}$ & $(5.1\pm0.4)\times10^{8}$ \\
$\beta$                                                   &  $1.61 \pm 0.03$                       &$1.48 \pm 0.09$&  $0.87 \pm 0.02$\\
$\epsilon_\mathrm{c}$ (MeV)                          &   $-$                                        & $70\pm 80$    &  $7$\\ 
$\chi^2_\mathrm{min}/${\it d.o.f.}           & 40.9/20 (0.38\%)                    &  40.6/19 (0.25\%) &43.8/20 (0.16\%)  \\ 
$d $ (m)                          & $300^{+390}_{-180}$& $250^{+430}_{-210}$& $100^{+280}$ \tablenotemark{a}\\ \hline
\end{tabular}
\\
\tablenotetext{a}{The lower value were unable to be determined.}
\end{table}
%
%
%
%
\begin{table}
 \caption{The $N_\mathrm{e}$ estimated for 071213 and 081225.}
 \label{tab:Ne}
\begin{tabular}{cccccc}
{} & \multicolumn{2}{c}{071213} & \multicolumn{2}{c}{081225} \\ \hline
$\theta_\mathrm{max} $(deg.)  &   $H=300$ m            & $H=1000$ m             &   $H=300$ m            & $H=1000$ m \\ 
15                                          &$1.9\times10^{10}$       &  $6.2\times10^{9}$   & $3.7\times 10^9$     & $1.2\times 10^9$ \\
30                                          &$4.7\times10^{11}$       &  $1.5\times10^{11}$ & $9.3\times10^{10}$  &$3.0\times10^{10}$  \\ \hline
\end{tabular}
\end{table}
\end{document}